\documentclass[final,5p,times]{elsarticle}
\pdfoutput=1
\usepackage[utf8]{inputenc}
\usepackage[T1]{fontenc}
\usepackage{amstext,amsmath,amssymb,amsfonts,bbm,euscript,color,dsfont}
\usepackage{epsfig}
\usepackage{slashed}
\usepackage{amsthm}
\usepackage{subfigure}
\usepackage{xcolor}
\usepackage{multirow}
\usepackage{bbold}
\usepackage{bm}
\usepackage{tikz}
\usetikzlibrary{positioning,arrows.meta}
\usepackage{verbatim}
\usepackage{graphicx}
\usepackage{latexsym}
\usepackage[colorlinks=true,citecolor=blue,linkcolor=blue,urlcolor=blue]{hyperref}
\newcommand{\Tr}{\operatorname{Tr}}
\newcommand{\FMR}{\mathrm{FMR}}

\journal{Physics Letters B}
\biboptions{sort&compress}
\begin{document}
\begin{frontmatter}
\title{A regulated zero-temperature construction of the Fundamental Modular Region in pure Yang--Mills theory and QCD}
\author[urjc]{Rodrigo Carmo Terin}
\ead{rodrigo.carmo@urjc.es}
\address[urjc]{King Juan Carlos University, Faculty of Experimental Sciences and Technology, Department of Applied Physics, Av.\ del Alcalde de M\'{o}stoles, 28933 Madrid, Spain}

\begin{abstract} 

We formulate the Fundamental Modular Region (FMR) as the zero-temperature limit of regulated copy-weighted Landau gauges. At finite $\beta$, the measure localizes on absolute minima, and the leading correction is dominated by the inverse Faddeev--Popov (FP) operator. A small $SU(2)$ benchmark provides a computational realization through population annealing and collective basin hopping. Our construction defines absolute Landau gauge without parametrizing the FMR boundary and extends naturally to full quantum chromodynamics (QCD). It also admits a Hamiltonian interpretation in terms of the gauge-fixed vacuum wave functional on the FMR. 

\end{abstract} 

\end{frontmatter}

\section{Introduction}

Yang-Mills (YM) theories furnish the gauge-theoretic backbone of the Standard Model and remain the paradigmatic framework for understanding non-Abelian quantum fields \cite{YangMills:1954ek}. Their perturbative quantization is usually formulated through the FP procedure \cite{FaddeevPopov:1967fc}, which is highly successful in the ultraviolet (UV) regime. In the infrared (IR), however, the same construction faces a fundamental obstruction: the gauge condition does not select a unique representative on each gauge orbit. Distinct gauge-equivalent configurations may satisfy the same gauge condition, giving rise to the Gribov ambiguity \cite{Gribov:1977wm,Singer:1978dk}; see \cite{Vandersickel:2012tz} for a review.

In the Landau gauge, infinitesimal Gribov copies are associated with zero modes of the FP operator
\begin{equation}
{\cal M}(A)=-\partial_\mu D_\mu(A).
\end{equation}
Gribov's proposal, later developed by Zwanziger, restricts the path integral to the first Gribov region $\Omega$, where ${\cal M}(A)$ is positive \cite{Gribov:1977wm,Zwanziger:1989mf,Zwanziger:1989wz}. The resulting Gribov--Zwanziger (GZ) and refined Gribov--Zwanziger (RGZ) approaches provide local and renormalizable descriptions of the restriction to the horizon and lead to IR gluon and ghost propagators in qualitative agreement with lattice simulations \cite{Dudal:2008sp,Dudal:2008rm,Dudal:2011gd,Cucchieri:2007rg,Dudal:2018cli}. In modern formulations, the use of the gauge-invariant transverse field $A_\mu^h$ also enables one to construct an exact nonperturbative BRST symmetry and to extend the framework beyond the Landau gauge \cite{Capri:2015ixa,Capri:2016gut,Capri:2017ab,Capri:2021}.
Yet the first Gribov region does not solve the full gauge-copy problem. It removes infinitesimal copies, but it still contains copies related by finite gauge transformations. The stronger geometric object is the FMR $\Lambda$, defined as the set of absolute minima of the Landau functional along each gauge orbit. Although every gauge orbit crosses the first Gribov region \cite{DellAntonio:1991xt}, the direct restriction of the path integral to $\Lambda$ is a global minimization problem. This is the central practical obstruction: $\Omega$ is characterized by the spectral condition ${\cal M}(A)>0$, whereas $\Lambda$ is defined by the variational condition
\begin{equation}
F_A[1]\leq F_A[U]\qquad \forall\, U.
\end{equation}
A local continuum implementation of this global condition is not known.
A different line of attack was proposed by J. Serreau, M. Tissier, and further collaborators, who introduced families of Landau gauges based on a weighted average over Gribov copies \cite{Serreau:2012cg,Serreau:2015yna,Tissier:2018mxa,Reinosa:2020mnx}. In these gauges, copies are weighted by a Boltzmann factor involving the Landau functional, schematically
\begin{equation}
\exp[-\beta F_A[U]],
\end{equation}
possibly supplemented by determinant factors controlled by a regulator $\zeta$. At finite $\beta$, this gives a smooth copy-weighted gauge fixing which admits a local replicated representation and is suitable for continuum calculations. This construction has usually been interpreted as a way of lifting the degeneracy among copies or favoring representatives closer to the absolute Landau gauge. In recent work, we used this viewpoint to build a unified ST--RGZ framework in which the horizon restriction and the copy-weighted sector coexist in a single local BRST-consistent formulation \cite{Terin:2026unified}. That analysis made explicit that the horizon sector sets the approach to the first Gribov region, whereas the weighted-copy sector controls the relative importance of Gribov copies on a fixed gauge orbit. A complementary development showed that, in a replica-broken branch, the ST determinant can also induce a leading nonlocal bilinear kernel with the same structure as the quadratic part of the BRST-invariant Gribov horizon functional \cite{terin2026emergentgribovhorizonreplica}. This work focuses on a different and more direct consequence of the copy-weighted sector: the zero-temperature limit $\beta\to\infty$ implements the absolute-minimum selection associated with the FMR.
The main observation is that the FMR can be defined operationally as the zero-temperature limit of copy-weighted Landau gauges. The parameter $\beta$ is naturally interpreted as an inverse temperature on the space of Gribov copies. At finite $\beta$, all copies contribute with different weights. In the limit $\beta\to\infty$, the orbitwise measure localizes on the absolute minima of the Landau functional. This gives the continuum counterpart of the best-copy prescription used in lattice Landau gauge:
\begin{equation}
\text{best copy}
\quad \longleftrightarrow \quad
\lim_{\beta\to\infty}
\text{copy-weighted gauge}.
\end{equation}
The result is not merely a formal steepest-descent statement. We define our construction with UV and IR regulators, for instance on a finite lattice with spacing $a$ and volume $V$, where the gauge group is compact and the Landau functional is continuous. At fixed regulator, the absolute minimum exists on every orbit and the $\beta\to\infty$ limit is an ordinary zero-temperature localization. The continuum FMR prescription is then attaines by an ordered sequence of limits.
A second output of our prescription is that the approach to the FMR is controlled. Around an isolated absolute representative $A^{U_*}$, the Landau functional expands quadratically with Hessian given by the FP operator ${\cal M}(A^{U_*})$. Consequently, the finite-$\beta$ correction to an orbit observable is regulated by the inverse FP operator,
\begin{equation}
\langle O\rangle_{\beta,[A]}
=
O[A^{U_*}]
+
\frac{1}{2\beta}
\Tr\left[
{\cal M}^{-1}(A^{U_*})\,O^{(2)}_{U_*}
\right]
+
O(\beta^{-2}).
\end{equation}
Thus the convergence toward the FMR is not arbitrary: it is governed by the low-lying spectrum of the FP operator at the absolute representative. This gives a concrete diagnostic for lattice tests, since configurations close to the FMR boundary are expected to exhibit the slowest convergence in $\beta$.
This perspective also clarifies the relation with horizon-based approaches. A GZ-type horizon term restricts the field measure toward the first Gribov region, but it does not distinguish between representatives on the same gauge orbit. The zero-temperature copy weight plays the complementary role of selecting the absolute minima along each orbit. In the $A^h$ formulation the horizon functional is gauge invariant and therefore constant on a fixed orbit; it changes the measure over gauge orbits, and at the same time the $\beta$-dependent Gibbs factor changes the relative weights of copies on a given orbit. The two mechanisms are therefore compatible but conceptually distinct:
\begin{equation}
\gamma\neq0:\quad \Omega,
\qquad
\beta\to\infty:\quad \Lambda.
\end{equation}
Our formulation is also directly connected with lattice practice. The distinction between minimal and absolute Landau gauge is approached numerically through best-copy or simulated-annealing procedures, which search for the extrema of the lattice Landau functional and reveal the dependence of IR propagators on Gribov copies \cite{Cucchieri:1997dx,Cucchieri:1997ns,Bornyakov:2008yx,Bogolubsky:2009dc,Sternbeck:2012mf}. Our present construction is not proposed as a novel optimization algorithm. Rather, it gives a field-theoretic interpretation of the zero-temperature/best-copy limit: finite $\beta$ defines a regulated copy-weighted gauge, and $\beta\to\infty$ gives the absolute Landau gauge/FMR selection. In this sense, simulated annealing is one possible numerical realization of the same localization principle, whereas the continuum construction provides an ordered limiting prescription and predicts how finite-$\beta$ observables should approach their best-copy values, with the leading correction setted by ${\cal M}^{-1}(A^{U_*})$.

The letter is organized as follows. In Sec.~\ref{sec:copy-weighted} we formulate copy-weighted Landau gauges as Gibbs measures on the set of Gribov copies of a fixed gauge orbit. In Sec.~\ref{sec:zero-temperature} we prove the zero-temperature localization on the absolute minima and derive the leading $1/\beta$ correction controlled by the inverse FP operator. In Sec.~\ref{sec:regulated-path-integral} we introduce the regulated FMR path-integral prescription and the corresponding ordered limit. In Sec.~\ref{sec:localized-version} we present the local finite-$\beta$ replicated representation, emphasizing that locality holds for the regulator and that the exact FMR projection is recovered only at $\beta=\infty$. In Sec.~\ref{sec:first-gribov-region} we clarify the relation with the first Gribov region and with horizon-based approaches. In Sec.~\ref{sec:lattice} we discuss the lattice realization in terms of best-copy and simulated-annealing procedures, together with the proposed convergence diagnostic. In Sec.~\ref{sec:computational-proof} we present a computational proof of concept based on population annealing, evolutionary neural proposals and collective basin hopping. In Sec.~\ref{sec:scope} we summarize the scope and possible extensions of our construction.
~\ref{app:regulated-localization} and \ref{app:finite-beta-saddle} provide further technical details on the regulated orbitwise localization and the finite-$\beta$ saddle expansion, respectively. Appendix~\ref{app:local-finite-beta} develops the local replicated formulation, while~\ref{app:horizon-compatibility} establishes its compatibility with the horizon restriction.~\ref{app:lattice-diagnostics} and \ref{app:computational-realization} contain the lattice diagnostics and computational implementation. ~\ref{sec:qcd-extension} extends our prescription to full QCD, ~\ref{app:hamiltonian-fmr} presents its Hamiltonian interpretation in terms of the vacuum wave functional on the FMR, and~\ref{app:ordered-limits} summarizes the complete ordered-limit prescription. 

\section{Copy-weighted Landau gauges} \label{sec:copy-weighted}

Let $A_\mu=A_\mu^aT^a$ be a YM gauge field. Along the gauge orbit of $A_\mu$, we define the Landau functional
\begin{equation} F_A[U] = \frac{1}{2} \int d^dx\,A_\mu^{U,a}A_\mu^{U,a}. 
\label{landau-functional} 
\end{equation} 
Its stationary points are Landau-gauge representatives, 
\begin{equation}
\frac{\delta F_A[U]}{\delta U}=0 \qquad \Longleftrightarrow \qquad \partial_\mu A_\mu^U=0. 
\label{stationary-landau} 
\end{equation} 
Different stationary points on the same gauge orbit are Gribov copies. The corresponding FP operator is 
\begin{equation} 
{\cal M}(A^U) = -\partial_\mu D_\mu(A^U). \label{fp-operator} 
\end{equation}
The first Gribov region is 
\begin{equation} \Omega = \left\{ A\,;\, \partial_\mu A_\mu=0, \quad {\cal M}(A)>0 \right\}, 
\label{omega-def}
\end{equation} 
whereas the FMR is the subset of absolute minima of $F_A[U]$ along each gauge orbit, \begin{equation}
\Lambda = \left\{ A\in\Omega\,;\, F_A[1]\leq F_A[U]\quad \forall\,U \right\}.
\label{fmr-def} 
\end{equation} 
The obstruction to a direct implementation of Eq.~\eqref{fmr-def} is that it requires a global minimization over all gauge transformations. We replace this direct geometric restriction by a Gibbs measure on the set of Landau copies. Let $U_i$ denote the Landau copies on a fixed gauge orbit and define \begin{equation}
F_i[A]\equiv F_A[U_i]. 
\label{Fi-def} 
\end{equation}
For a gauge-dependent observable $O$, we introduce the orbitwise copy average \begin{equation}
\langle O\rangle_{\beta,[A]} = \frac{ \sum_i O[A^{U_i}]\,\exp[-\beta F_i[A]] }{ \sum_i \exp[-\beta F_i[A]] }. 
\label{copy-average-beta} 
\end{equation} 
The parameter $\beta$ plays the role of an inverse temperature on the space of Gribov copies. At finite $\beta$, all copies contribute with different weights. The FMR limit is obtained by sending $\beta$ to infinity. A more general copy-weighted average may include the determinant prefactor used in the Serreau--Tissier (ST) construction, 
\begin{equation}
\begin{aligned}
W_\zeta[A,U_i]
&=
s(i)
\frac{
\det\left({\cal M}[A^{U_i}]+\zeta{\bf 1}\right)
}{
|\det{\cal M}[A^{U_i}]|
},
\\
s(i)
&=
\mathrm{sign}\det{\cal M}[A^{U_i}].
\end{aligned}
\label{det-prefactor}
\end{equation}
where $\zeta>0$ regularizes possible zero modes of the FP operator. The corresponding copy average is
\begin{equation}
\langle O\rangle_{\beta,\zeta,[A]} = \frac{ \sum_i W_\zeta[A,U_i]\, O[A^{U_i}]\, \exp[-\beta F_i[A]] }{ \sum_i W_\zeta[A,U_i]\, \exp[-\beta F_i[A]] }. 
\label{copy-average-beta-zeta}
\end{equation}
The exponential factor controls the absolute-minimum selection, while $W_\zeta$ affects only the residual weighting among copies that remain degenerate in the zero-temperature limit.

\section{Zero-temperature localization and controlled approach to the FMR} 
\label{sec:zero-temperature}
Let \begin{eqnarray} 
F_*[A] &=& \min_i F_i[A], \nonumber\\ {\cal I}_*[A] &=& \left\{i\,;\,F_i[A]=F_*[A]\right\} \label{minimum-set} 
\end{eqnarray} 
be the absolute minimum of the Landau functional on the orbit and the set of copies realizing it. Writing 
\begin{equation} F_i[A]=F_*[A]+\Delta_i[A], \qquad \Delta_i[A]\geq0, 
\label{delta-def}
\end{equation} 
Eq.~\eqref{copy-average-beta} becomes \begin{equation} 
\langle O\rangle_{\beta,[A]} = \frac{ \sum_i O[A^{U_i}]\,\exp[-\beta\Delta_i[A]] }{ \sum_i \exp[-\beta\Delta_i[A]] }.
\label{copy-average-delta} 
\end{equation} 
All terms with $\Delta_i[A]>0$ are exponentially suppressed when $\beta\to\infty$. Hence, for any bounded orbit observable, \begin{equation}
\lim_{\beta\to\infty} \langle O\rangle_{\beta,[A]} = \frac{ \sum_{i\in{\cal I}_*[A]}O[A^{U_i}] }{ |{\cal I}_*[A]| }. \label{zero-temperature-result} 
\end{equation}
If the absolute minimum is unique, ${\cal I}_*[A]=\{i_*\}$, this reduces to 
\begin{equation} 
\lim_{\beta\to\infty} \langle O\rangle_{\beta,[A]} = O[A^{U_{i_*}}], \qquad F_A[U_{i_*}] = \min_U F_A[U]. 
\label{unique-minimum-result} 
\end{equation}
Thus $A^{U_{i_*}}$ belongs to the FMR. If several absolute minima exist, the limiting prescription gives an average over the degenerate representatives. These degeneracies correspond to possible boundary identifications of the FMR and are automatically retained by the limiting measure. With the determinant prefactor included, one similarly obtains \begin{equation} 
\lim_{\beta\to\infty} \langle O\rangle_{\beta,\zeta,[A]} = \frac{ \sum_{i\in{\cal I}_*[A]} W_\zeta[A,U_i]\,O[A^{U_i}] }{ \sum_{i\in{\cal I}_*[A]} W_\zeta[A,U_i] },
\label{weighted-zero-temp-result} \end{equation} 
provided that the denominator does not vanish on the minimizing set. Therefore, determinant factors may change the residual weighting among degenerate absolute minima, but they do not change the zero-temperature support of the measure. The FMR selection is driven by the Gibbs factor $\exp[-\beta F_i]$. The same statement can be written through the soft-min identity 
\begin{equation} \min_U F_A[U] = -\lim_{\beta\to\infty} \frac{1}{\beta} \ln \int{\cal D}U\, \exp[-\beta F_A[U]]. \label{soft-min} 
\end{equation}
Thus the FMR condition is generated dynamically by zero-temperature localization on the gauge orbit, rather than by an explicit parametrization of the boundary of $\Lambda$. Beyond the strict zero-temperature limit, the construction also gives a controlled approach to the FMR. Let $U_*$ be an isolated absolute minimum of $F_A[U]$ on a given orbit. Writing a nearby gauge transformation as 
\begin{equation}
U=U_*\exp(ig\theta), 
\label{nearby-transformation} 
\end{equation} 
one has, to quadratic order, 
\begin{equation} F_A[U] = F_A[U_*] + \frac{1}{2} \int d^dx\,d^dy\, \theta^a(x)\, {\cal M}^{ab}(A^{U_*};x,y)\, \theta^b(y) + O(\theta^3), 
\label{quadratic-expansion-around-fmr} \end{equation} 
where ${\cal M}(A^{U_*})$ is the FP operator evaluated at the absolute representative. The finite-$\beta$ average can then be expanded by the Laplace method. For a smooth orbit observable $O[A^U]$, one obtains \begin{equation} 
\langle O\rangle_{\beta,[A]} = O[A^{U_*}] + \frac{1}{2\beta} \Tr\left[ {\cal M}^{-1}(A^{U_*})\,O^{(2)}_{U_*} \right] + O(\beta^{-2}), 
\label{one-over-beta-expansion} 
\end{equation} 
in which $O^{(2)}_{U_*}$ denotes the second variation of $O[A^U]$ with respect to the orbit coordinates $\theta$ at $U=U_*$. The leading correction to the FMR limit is therefore governed by the inverse FP operator at the absolute representative. This makes the approach to the FMR calculable at finite $\beta$ and distinguishes the construction from a purely formal statement about the existence of absolute minima.

\section{Regulated FMR path integral} \label{sec:regulated-path-integral} 

The orbitwise construction induces a regulated path-integral prescription. We first introduce an UV and IR regulator. For instance, consider the theory on a finite lattice with spacing $a$ and finite volume $V$. The gauge group is then a compact product of $SU(N)$ groups, and the lattice Landau functional 
\begin{equation} {\cal F}_U[g] = -\sum_{x,\mu} \mathrm{Re\,Tr}\, g_x U_{x,\mu}g^\dagger_{x+\hat\mu} 
\label{lattice-functional-main} 
\end{equation} 
is continuous on a compact space. Hence, for every lattice gauge orbit, an absolute minimum exists. At fixed $(a,V)$, introduce the orbitwise normalization
\begin{equation}
{\cal N}_{\beta,\zeta}[U_{\rm lat}]
=
\int{\cal D}g\,
W_\zeta[U_{\rm lat},g]\,
e^{-\beta{\cal F}_{U_{\rm lat}}[g]}.
\label{orbit-normalization-main}
\end{equation}
and define
\begin{eqnarray}
\langle O\rangle_{\beta,\zeta}^{a,V}
&=&
\frac{1}{Z_{\rm YM}^{a,V}}
\int{\cal D}U_{\rm lat}\,
e^{-S_{\rm YM}[U_{\rm lat}]}
\nonumber\\
&&\times
\frac{
\int{\cal D}g\,
W_\zeta[U_{\rm lat},g]\,
O[U_{\rm lat}^g]\,
e^{-\beta{\cal F}_{U_{\rm lat}}[g]}
}{
{\cal N}_{\beta,\zeta}[U_{\rm lat}]
},
\label{regulated-lattice-expectation}
\end{eqnarray}
with
\begin{equation}
Z_{\rm YM}^{a,V}
=
\int{\cal D}U_{\rm lat}\,
e^{-S_{\rm YM}[U_{\rm lat}]}.
\label{ym-partition-main}
\end{equation} 
The FMR expectation value is then defined by the ordered limit 
\begin{equation} 
\boxed{ \langle O\rangle_{\rm FMR} = \lim_{a\to0} \lim_{V\to\infty} \lim_{\zeta\to0^+} \lim_{\beta\to\infty} \langle O\rangle_{\beta,\zeta}^{a,V}. } \label{regulated-fmr-limit-main} 
\end{equation} 
The order of limits is part of the definition. First, $\beta\to\infty$ performs the absolute-minimum selection at fixed regulator. Then $\zeta\to0^+$ removes the FP regulator. Finally, the infinite-volume and continuum limits are taken. The formal continuum counterpart of Eq.~\eqref{regulated-lattice-expectation} is
\begin{eqnarray}
\langle O\rangle_{\beta,\zeta}
&=&
\frac{1}{Z_{\rm YM}}
\int{\cal D}A\,e^{-S_{\rm YM}[A]}
\nonumber\\
&&\times
\frac{
\sum_i W_\zeta[A,U_i]\,O[A^{U_i}]\,e^{-\beta F_A[U_i]}
}{
\sum_i W_\zeta[A,U_i]\,e^{-\beta F_A[U_i]}
}.
\label{continuum-regulated-formal}
\end{eqnarray}
Equation~\eqref{continuum-regulated-formal} should be understood as shorthand for the regulated definition \eqref{regulated-fmr-limit-main}. The genuine definition is the ordered limit at fixed regulator, where the existence of absolute minima is guaranteed. The construction therefore solves the implementation problem in an operational sense. The FMR is not introduced by giving a global parametrization of its boundary. Instead, it is attained as the zero-temperature limit of a regulated copy-weighted gauge. The price one pays is that the exact FMR projection is global, as it must be: it depends on the absolute minimum of the Landau functional on the whole orbit. 

\section{Local finite-\texorpdfstring{$\beta$}{beta} representation} 
\label{sec:localized-version} 

The finite-$\beta$ regulator admits a local continuum representation. This is the local field theory associated with the copy-weighted gauge before the zero-temperature limit is taken. It should be distinguished from the exact FMR projection at $\beta=\infty$, which remains a global operation. The ST copy average can be localized by introducing a group-valued field and using the replica trick \cite{ParisiSourlas1979} to represent the denominator of the copy average. A convenient superspace representation uses nonlinear sigma-model superfields 
\begin{equation} {\cal V}_k(x,\theta,\bar\theta)\in SU(N), \qquad k=2,\ldots,n, 
\label{replica-superfields}
\end{equation} with
\begin{equation}
{\cal V}_k^\dagger{\cal V}_k={\bf 1}. \label{unitarity-replica-superfields} \end{equation}
The covariant derivative is 
\begin{equation} {\cal D}_\mu{\cal V}_k = \partial_\mu{\cal V}_k + ig\,{\cal V}_k A_\mu. \label{replica-covariant-derivative} \end{equation}
The local finite-$\beta$ action is \begin{eqnarray}
S_{\rm loc}^{(\beta,\zeta)} &=& S_{\rm YM}[A] + S_{\rm FP}[A,c,\bar c,b] + \sum_{k=2}^{n} S_{\rm ST}[A,{\cal V}_k;\beta,\zeta],\nonumber\\
\label{local-st-action}
\end{eqnarray}
in which
\begin{eqnarray}
S_{\rm ST}[A,{\cal V}_k;\beta,\zeta]
&=&
\frac{1}{g^2}
\int d^dx\,d\theta\,d\bar\theta\,
\mathrm{tr}\Big[
({\cal D}_\mu{\cal V}_k)^\dagger
({\cal D}_\mu{\cal V}_k)
\nonumber\\
&&\hspace{2.1cm}
+2\zeta\,\bar\theta\theta\,
\partial_{\bar\theta}{\cal V}_k^\dagger
\partial_\theta{\cal V}_k
\Big]
\nonumber\\[-0.7ex]
&&
\label{local-st-super}
\end{eqnarray} 
The singled replica gives the usual massive FP/Curci--Ferrari (CF) type sector, \begin{equation}
S_{\rm gf}^{(\beta,\zeta)} = \int d^dx \left[ \partial_\mu\bar c^aD_\mu^{ab}(A)c^b + \zeta\,\bar c^ac^a + ib^a\partial_\mu A_\mu^a + \frac{\beta}{2}A_\mu^aA_\mu^a \right]. \label{component-st-local}
\end{equation}
The denominator of the copy average is implemented by the replica limit, 
\begin{equation} 
\langle O\rangle_{\beta,\zeta} = \lim_{n\to0} \frac{ \int{\cal D}\Phi\, O[A]\, \exp[-S_{\rm loc}^{(\beta,\zeta)}] }{ \int{\cal D}\Phi\, \exp[-S_{\rm loc}^{(\beta,\zeta)}] }, \label{replica-limit-local}
\end{equation} 
where ${\cal D}\Phi$ denotes integration over the gauge, ghost, multiplier and replica fields. At finite $\beta$, Eqs.~\eqref{local-st-action}--\eqref{replica-limit-local} define a local field theory. This local theory is the regulator of the FMR prescription. The FMR limit is obtained only after the local finite-$\beta$ theory has been defined: 
\begin{equation} 
\langle O\rangle_{\rm FMR} = \lim_{a\to0} \lim_{V\to\infty} \lim_{\zeta\to0^+} \lim_{\beta\to\infty} \lim_{n\to0} \langle O\rangle_{n,\beta,\zeta}^{a,V}. \label{localized-fmr-limit} 
\end{equation} 
The strict FMR projection is therefore not represented by a simple local action at $\beta=\infty$; rather, it is attained as the zero-temperature limit of the local finite-$\beta$ family. If one wants to keep the relation with the first Gribov region explicit, one may add the BRST-invariant RGZ block written in terms of $A_\mu^h$. The local finite-$\beta$ action then becomes \begin{eqnarray}
S_{\rm loc}^{(\beta,\zeta,\gamma)} &=& S_{\rm YM} + S_{\rm FP} + \sum_{k=2}^{n} S_{\rm ST}[A,{\cal V}_k;\beta,\zeta] \nonumber\\ && - \int d^dx \left[ \bar\varphi_\mu^{ac}{\cal M}^{ab}(A^h)\varphi_\mu^{bc} - \bar\omega_\mu^{ac}{\cal M}^{ab}(A^h)\omega_\mu^{bc} \right] \nonumber\\ && - g\gamma^2 f^{abc} \int d^dx\, (A^h)_\mu^a (\varphi+\bar\varphi)_\mu^{bc} + S_{\rm transv}[A^h]\nonumber\\&&+ S_{\rm ref}[A^h,\varphi,\omega]. 
\label{local-st-rgz-action}
\end{eqnarray}
Here $S_{\rm transv}$ represents $\partial_\mu A_\mu^h=0$, and $S_{\rm ref}$ denotes the usual RGZ refinement terms proportional to $(A^h)^2$ and $\bar\varphi\varphi-\bar\omega\omega$. This localized action makes the separation of mechanisms explicit:
\begin{equation}
\gamma\neq0 \quad\Longrightarrow\quad \text{restriction toward the first Gribov region}, 
\label{gamma-omega}
\end{equation} 
whereas
\begin{equation} 
\beta\to\infty \quad\Longrightarrow\quad \text{absolute-minimum selection on each orbit}.
\label{beta-lambda}
\end{equation} 
Thus the horizon sector controls the restriction to $\Omega$, and the zero-temperature copy-weighted sector governs the localization toward $\Lambda$.

\section{Relation with the first Gribov region} \label{sec:first-gribov-region} 

The first Gribov region and the FMR implement two different layers of gauge fixing. The first Gribov region is defined by the spectral condition
\begin{equation}
A\in\Omega \qquad \Longleftrightarrow \qquad \partial_\mu A_\mu=0, \qquad {\cal M}(A)>0, \label{omega-condition-new}
\end{equation} 
whereas the FMR is defined by the global variational condition
\begin{equation} 
A\in\Lambda \qquad \Longleftrightarrow \qquad A\in\Omega, \qquad F_A[1]\leq F_A[U]\quad\forall U. 
\label{lambda-condition-new} 
\end{equation} 
Therefore, 
\begin{equation} 
\Lambda\subset\Omega. 
\label{lambda-inside-omega-new}
\end{equation} 
The positivity of ${\cal M}(A)$ removes infinitesimal copies associated with zero modes of the FP operator, but it does not select the absolute minimum of the Landau functional on the full gauge orbit. Thus a restriction to $\Omega$ still leaves residual finite Gribov copies. The zero-temperature prescription addresses precisely this remaining step. The Gibbs factor $\exp[-\beta F_A[U]]$ suppresses all nonminimal representatives and localizes the orbitwise measure on the absolute minima. This gives the hierarchy
\begin{equation}
\text{Landau gauge} \quad\longrightarrow\quad \Omega \quad\longrightarrow\quad \Lambda, \label{hierarchy-new}
\end{equation} 
where the first nontrivial restriction is setted by the positivity of the FP operator, and the second is governed by absolute minimization of the Landau functional. This distinction also clarifies the relation with horizon-based approaches. A GZ horizon term implements the restriction to the first Gribov region. It does not, by itself, distinguish between representatives on the same gauge orbit. The copy-weighted zero-temperature limit plays a different role: it selects, or averages over, the absolute minima of the orbit. The two mechanisms are complementary rather than identical. In particular, our present construction should not be interpreted as a novel local horizon functional for the FMR. It is a local finite-$\beta$ regulator whose zero-temperature limit implements the FMR selection. The two operations are compatible at the orbitwise level. Since the horizon functional in the $A^h$ formulation is gauge invariant, it is constant along a fixed gauge orbit,
\begin{equation} 
H(A^{h,U_i})=H(A^h). 
\label{horizon-constant-orbit}
\end{equation} 
Therefore, in a combined ST--RGZ weight, \begin{equation}
\exp[-\beta F_A[U_i]-\gamma^4H(A^{h,U_i})], \label{combined-weight}
\end{equation}
the horizon factor cancels in the normalized copy average on each orbit and does not modify the relative copy weights. The $\gamma$-sector restricts the field measure toward $\Omega$, and the $\beta$-sector performs the absolute-minimum selection among copies. Hence the two limits do not compete at the level of copy selection: 
\begin{equation} 
\lim_{\beta\to\infty} \frac{ \sum_i O[A^{U_i}] e^{-\beta F_A[U_i]-\gamma^4H(A^h)} }{ \sum_i e^{-\beta F_A[U_i]-\gamma^4H(A^h)} } = \lim_{\beta\to\infty} \langle O\rangle_{\beta,[A]}. 
\label{gamma-cancels-copy-average}
\end{equation} 
This is the precise sense in which the horizon restriction and the FMR localization are compatible: $\gamma$ changes the measure over gauge orbits, while $\beta$ changes the relative weight of representatives on a fixed orbit. 

\section{Lattice realization and tests} \label{sec:lattice} 

Our approach has a direct lattice realization. The finite-$\beta$ measure
\begin{equation} 
P_\beta[g|U] = \frac{ \exp[-\beta{\cal F}_U[g]] }{ \int{\cal D}g\,\exp[-\beta{\cal F}_U[g]] } \label{lattice-beta-measure} 
\end{equation} 
is a finite-temperature version of lattice Landau gauge fixing. The limit $\beta\to\infty$ gives the best-copy prescription:
\begin{equation} 
\lim_{\beta\to\infty}P_\beta[g|U] \quad \text{is supported on} \quad g_*=\arg\min_g{\cal F}_U[g]. 
\label{best-copy-limit-new}
\end{equation}
Thus simulated annealing and best-copy algorithms are lattice implementations of the same zero-temperature localization principle. The role of the present construction is not to prescribe a particular optimizer or to remove the intrinsic difficulty of locating the global extremum. It is to define the corresponding field-theoretic limit and to provide analytic control of the approach to it. This gives a concrete test of our formulation. One may compute gauge-dependent correlators in three ensembles: 
\begin{equation}
\begin{aligned}
&
\text{ordinary Landau gauge},
\qquad
\text{finite-}\beta\text{ copy-weighted gauge},
\\
&
\text{best-copy gauge}.
\end{aligned}
\end{equation}
For example, one can monitor the gluon and ghost propagators 
\begin{equation}
D_\beta(p), \qquad G_\beta(p), \label{beta-dependent-propagators} 
\end{equation}
and study whether the limits
\begin{equation}
D_{\rm FMR}(p) = \lim_{\beta\to\infty}D_\beta(p), \qquad G_{\rm FMR}(p) = \lim_{\beta\to\infty}G_\beta(p)
\label{fmr-propagator-limits}
\end{equation} 
are stable when the number of gauge copies, the lattice volume and the lattice spacing are varied. The $1/\beta$ expansion in Eq.~\eqref{one-over-beta-expansion} also gives a practical diagnostic. For each orbit, the leading deviation from the best-copy value is controlled by the inverse FP operator at the absolute representative. Therefore, configurations close to the FMR boundary, where small FP eigenvalues appear, are expected to show the slowest convergence in $\beta$. This gives an explicit and testable prediction of the copy-weighted approach: 
\begin{equation}
\begin{aligned}
\delta O_\beta
&\equiv
\langle O\rangle_{\beta,[A]}-O[A^{U_*}]
\\
&=
O(\beta^{-1}),
\qquad
\text{with coefficient governed by }
{\cal M}^{-1}(A^{U_*}).
\end{aligned}
\label{beta-convergence-diagnostic}
\end{equation}
Thus the finite-$\beta$ approach to the FMR can be tested by both extrapolating correlators, and by correlating the convergence rate with the lowest eigenvalues of the FP operator at the best copy.

\section{Computational proof of concept}
\label{sec:computational-proof}

Our regulated prescription can be implemented directly on the gauge orbit.
To demonstrate this point, we carried out a small $SU(2)$ proof of concept
combining local overrelaxation, simulated and population annealing, a
low-dimensional neural warm start optimized by an evolutionary strategy, and
collective basin-hopping moves. Evolutionary and neural-assisted gauge fixing
have been explored previously in different settings
\cite{OliveiraSilva:2001evolution,Hsiao:2026mlgauge}; here they are used only as
proposal mechanisms for the finite-$\beta$ measure defined above. Every
candidate is subsequently relaxed to the lattice Landau condition, so that the
final comparison is made between stationary representatives on the same gauge
orbit.

The population-annealing implementation samples a finite sequence of Gibbs
measures $P_{\beta_k}[g|U]$ by Boltzmann reweighting, resampling and local
Metropolis sweeps \cite{Machta:2010population}. In addition, the
$1/\beta$ analysis suggests a collective proposal based on the low-lying FP
modes of a locally converged copy,
\begin{equation}
 {\cal M}(U^{g_*})\psi_n=\lambda_n\psi_n,
 \qquad
 h(x)=\exp\left[
 i\sum_{n=1}^{N_{\rm low}}
 \frac{\xi_n}{\sqrt{\lambda_n+\mu^2}}
 \psi_n^a(x)T^a
 \right],
\label{fp-guided-proposal-main}
\end{equation}
where the $\xi_n$ are Gaussian variables and the overall amplitude is scanned.
After applying $h$, ordinary overrelaxation maps the candidate to a new local
extremum. Equation~\eqref{fp-guided-proposal-main} therefore turns the FP
spectral information into an explicit basin-hopping move.

We tested the construction on twelve configurations from an exploratory
$SU(2)$ Wilson-action ensemble on a $4^4$ lattice at $\beta_{\rm W}=1.2$.
This ensemble exhibits several distinct extrema under
multistart gauge fixing. Four random starts were used for the low-budget
methods. The collective methods used twenty-four proposals, and a
thirty-two-start overrelaxation search was included as a reference. The
``hit rate'' in Table~\ref{tab:computational-fmr-benchmark} refers only to the
lowest value found by any method in this finite benchmark; it is not a proof
that the absolute minimum has been identified.

\begin{table*}[t]
\centering
\caption{Proof-of-concept gauge-fixing benchmark on twelve configurations from an exploratory $SU(2)$ Wilson-action ensemble on a $4^4$ lattice at $\beta_{\rm W}=1.2$. The gap is measured relative to the lowest value found by any method for each orbit, so the hit rate is not a certificate of the exact FMR representative. Times from the unoptimized prototype are not used in the comparison.}
\label{tab:computational-fmr-benchmark}
\begin{tabular}{lccc}
\hline\hline
Method & Mean gap & Hit rate ($10^{-7}$) & Mean Landau residual \\
\hline
Four-start overrelaxation & $6.12\times10^{-6}$ & $92\%$ & $1.20\times10^{-8}$ \\
Neural warm start $+$ overrelaxation & $5.20\times10^{-5}$ & $92\%$ & $1.09\times10^{-8}$ \\
Simulated annealing $+$ overrelaxation & $1.79\times10^{-5}$ & $58\%$ & $2.79\times10^{-6}$ \\
Population annealing & $1.11\times10^{-8}$ & $100\%$ & $5.47\times10^{-8}$ \\
Isotropic collective hops & $6.49\times10^{-11}$ & $100\%$ & $1.14\times10^{-10}$ \\
FP-guided collective hops & $3.29\times10^{-11}$ & $100\%$ & $7.42\times10^{-11}$ \\
Neural warm start $+$ FP hops & $1.25\times10^{-10}$ & $100\%$ & $1.93\times10^{-10}$ \\
Thirty-two-start overrelaxation reference & $6.12\times10^{-12}$ & $100\%$ & $3.37\times10^{-11}$ \\
\hline\hline
\end{tabular}
\end{table*}

The calculation shows that our zero-temperature prescription is computationally
constructive rather than merely formal. On the most clearly trapped orbit, the
four-start overrelaxation result remained approximately $7.3\times10^{-5}$
above the lowest value found, whereas population annealing and both collective
hop constructions recovered the lower basin. Over the twelve configurations,
population annealing and the collective methods reached the benchmark-best
value within $10^{-7}$ on every orbit. Details, per-configuration
results and additional figures are provided in the appendices.

\section{Conclusion and outlook} \label{sec:scope} 

We have reformulated the restriction to the FMR as a zero-temperature localization problem. At finite regulator, the gauge group is compact and the Landau functional has absolute minima on every orbit. The copy-weighted measure 
\begin{equation}
P_\beta[g|U]\propto e^{-\beta{\cal F}_U[g]}
\end{equation} 
therefore has a well-defined $\beta\to\infty$ limit supported on the absolute minima. The continuum FMR is defined by the ordered limit \begin{equation} 
\langle O\rangle_{\rm FMR} = \lim_{a\to0} \lim_{V\to\infty} \lim_{\zeta\to0^+} \lim_{\beta\to\infty} \langle O\rangle_{\beta,\zeta}^{a,V}. 
\end{equation} 
Our construction resolves the practical implementation problem in an operational sense. It does not require an explicit parametrization of the boundary of $\Lambda$. Instead, the FMR is obtained as the zero-temperature limit of a regulated family of copy-weighted gauges. At finite $\beta$, this family admits a local replicated representation. At $\beta=\infty$, the exact projection is necessarily global, because it represents an absolute minimization over the full gauge orbit. A central output of the construction is that the approach to the FMR is controlled. For an isolated absolute minimum, the leading finite-$\beta$ correction is governed by
\begin{equation}
\frac{1}{2\beta} \Tr\left[ {\cal M}^{-1}(A^{U_*})\,O^{(2)}_{U_*} \right],
\end{equation}
showing that the inverse FP operator controls the rate at which the copy-weighted gauge approaches the absolute representative. This gives a calculable correction to the FMR projection and a concrete diagnostic for lattice tests. Our result also clarifies the relation between the FMR, the first Gribov region and horizon-based approaches. The horizon condition restricts the theory to $\Omega$, and the zero-temperature copy weight selects the absolute minima inside each orbit. These are different and complementary operations: 

\begin{equation}
\gamma\neq0:\quad \Omega, \qquad \beta\to\infty:\quad \Lambda.
\end{equation} 
Thus our present prescription turns the FMR restriction into a concrete localization problem, accessible analytically through the finite-$\beta$ local regulator and numerically through copy-weighted sampling and global-search procedures. The exploratory $SU(2)$ benchmark shows explicitly that population annealing and collective basin-hopping moves can recover lower extrema missed by a low-budget multistart search. The same ordered prescription extends to full QCD by including the quark action before the zero-temperature copy-selection limit is taken. Since the quark action is gauge invariant, the orbitwise selection remains setted by the Landau functional, and the limit $\beta\to\infty$ defines QCD Green functions in absolute Landau gauge, or equivalently in the FMR prescription. This does not constitute a closed-form solution of QCD, but it furnishes a regulated definition of the gauge-fixed nonperturbative QCD functional integral after the residual Landau-gauge Gribov ambiguity has been removed. Gauge-invariant observables are unchanged by the copy average, and gauge-dependent IR quantities, e.g. the gluon, ghost and quark propagators, acquire a well-defined FMR prescription.

Our construction also admits a Hamiltonian interpretation, developed in the appendices. In the Schr\"odinger representation, physical vacuum wave functionals are constant along gauge orbits as a consequence of Gauss' law, whereas gauge-dependent vacuum correlators require a choice of representative on each orbit. The zero-temperature copy-weighted prescription provides this choice by selecting the absolute Landau representative, thereby giving a regulated representation of the vacuum wave functional on the FMR. This does not amount to solving the YM or QCD Hamiltonian explicitly, but it clarifies how the gauge-fixed nonperturbative vacuum can be represented after the residual Landau-gauge Gribov ambiguity has been removed.
Future work should focus on the stability of the ordered limit for IR Green functions. In particular, the behavior of $D_\beta(p)$ and $G_\beta(p)$ as $\beta$, the number of copies, the lattice volume and the lattice spacing are varied provides a direct test of whether the zero-temperature prescription leads to a universal FMR limit. The predicted relation between the $1/\beta$ convergence rate and the low-lying spectrum of the FP operator at the best copy gives a sharper observable target for such tests.

\clearpage
\appendix
\numberwithin{equation}{section}
\numberwithin{figure}{section}
\numberwithin{table}{section}
\section{Regulated orbitwise localization}
\label{app:regulated-localization}

At fixed ultraviolet and infrared regulator the orbitwise minimization problem is ordinary and well defined.
For definiteness, consider a finite lattice with spacing $a$ and finite volume $V$.
The gauge group on the lattice is the compact product
\begin{equation}
{\cal G}_{a,V}
=
\prod_x SU(N),
\label{app:gauge-group-compact}
\end{equation}
and the lattice Landau functional
\begin{equation}
{\cal F}_U[g]
=
-\sum_{x,\mu}
{\rm Re\,Tr}\,
g_x U_{x,\mu}g^\dagger_{x+\hat\mu}
\label{app:lattice-landau-functional}
\end{equation}
is a continuous function on ${\cal G}_{a,V}$.
Therefore, by compactness, for every lattice gauge orbit there exists at least one absolute minimum,
\begin{equation}
{\cal F}_*
=
\min_{g\in{\cal G}_{a,V}}
{\cal F}_U[g].
\label{app:absolute-minimum-exists}
\end{equation}
The lattice FMR is the set of representatives satisfying
\begin{equation}
{\cal F}_U[g_*]
=
{\cal F}_*.
\label{app:lattice-fmr-definition}
\end{equation}

We define the positive copy-weighted measure
\begin{equation}
P_\beta[g|U]
=
\frac{
\exp[-\beta{\cal F}_U[g]]
}{
Z_\beta[U]
},
\qquad
Z_\beta[U]
=
\int_{{\cal G}_{a,V}}{\cal D}g\,
\exp[-\beta{\cal F}_U[g]].
\label{app:positive-copy-measure}
\end{equation}
Equivalently, subtracting the absolute minimum,
\begin{equation}
{\cal F}_U[g]
=
{\cal F}_*
+
\Delta_U[g],
\qquad
\Delta_U[g]\geq0,
\label{app:delta-definition}
\end{equation}
we get
\begin{equation}
P_\beta[g|U]
=
\frac{
\exp[-\beta\Delta_U[g]]
}{
\int_{{\cal G}_{a,V}}{\cal D}g\,
\exp[-\beta\Delta_U[g]]
}.
\label{app:positive-copy-measure-delta}
\end{equation}

Let ${\cal M}_*$ denote the set of absolute minima,
\begin{equation}
{\cal M}_*
=
\left\{
g\in{\cal G}_{a,V}\,;\,
\Delta_U[g]=0
\right\}.
\label{app:minimizing-set}
\end{equation}
For any continuous bounded function ${\cal O}[U^g]$ on the orbit, the finite-$\beta$ copy average is
\begin{equation}
\langle{\cal O}\rangle_{\beta,[U]}
=
\int_{{\cal G}_{a,V}}{\cal D}g\,
P_\beta[g|U]\,
{\cal O}[U^g].
\label{app:finite-beta-copy-average}
\end{equation}
In the zero-temperature limit the measure localizes on ${\cal M}_*$.
More explicitly, for every open neighborhood ${\cal N}$ of ${\cal M}_*$, compactness implies that
\begin{equation}
\Delta_{\cal N}
=
\inf_{g\notin{\cal N}}\Delta_U[g]
>
0.
\label{app:gap-away-from-minima}
\end{equation}
Thus
\begin{equation}
\int_{{\cal G}_{a,V}\setminus{\cal N}}{\cal D}g\,
P_\beta[g|U]
\leq
C_{\cal N}\,
\exp[-\beta\Delta_{\cal N}],
\label{app:exponential-suppression-away}
\end{equation}
for some regulator-dependent constant $C_{\cal N}$.
Hence
\begin{equation}
\lim_{\beta\to\infty}
\int_{{\cal G}_{a,V}\setminus{\cal N}}{\cal D}g\,
P_\beta[g|U]
=
0,
\label{app:weak-support-minima}
\end{equation}
which proves that every weak limit of $P_\beta[g|U]$ is supported on ${\cal M}_*$.

If the absolute minimum is unique, ${\cal M}_*=\{g_*\}$, then
\begin{equation}
\lim_{\beta\to\infty}
P_\beta[g|U]
=
\delta[g-g_*],
\label{app:unique-delta-limit}
\end{equation}
in the weak sense, and therefore
\begin{equation}
\lim_{\beta\to\infty}
\langle{\cal O}\rangle_{\beta,[U]}
=
{\cal O}[U^{g_*}].
\label{app:unique-observable-limit}
\end{equation}
If several absolute minima exist, the limiting measure is supported on the degenerate minimizing set.
In that case, the zero-temperature prescription automatically retains the residual average over degenerate representatives.
These degeneracies are the lattice counterpart of possible boundary identifications of the continuum FMR.

The same reasoning applies to the continuum notation used in the main text.
Let $U_i$ be the Landau copies on a fixed orbit and
\begin{eqnarray}
F_i[A]
&=&
F_A[U_i],
\nonumber\\
F_*[A]
&=&
\min_i F_i[A],
\nonumber\\
{\cal I}_*[A]
&=&
\{i\,;\,F_i[A]=F_*[A]\}
\nonumber\\[-0.7ex]
&&
\label{app:continuum-minima}
\end{eqnarray}
Then
\begin{equation}
\langle O\rangle_{\beta,[A]}
=
\frac{
\sum_i O[A^{U_i}]\exp[-\beta F_i[A]]
}{
\sum_i \exp[-\beta F_i[A]]
}
=
\frac{
\sum_i O[A^{U_i}]\exp[-\beta\Delta_i[A]]
}{
\sum_i \exp[-\beta\Delta_i[A]]
},
\label{app:continuum-copy-average-delta}
\end{equation}
where
\begin{equation}
\Delta_i[A]=F_i[A]-F_*[A]\geq0.
\label{app:continuum-delta}
\end{equation}
Consequently,
\begin{equation}
\lim_{\beta\to\infty}
\langle O\rangle_{\beta,[A]}
=
\frac{
\sum_{i\in{\cal I}_*[A]}O[A^{U_i}]
}{
|{\cal I}_*[A]|
}.
\label{app:continuum-zero-temp-result}
\end{equation}

If determinant prefactors are included, one obtains instead
\begin{equation}
\langle O\rangle_{\beta,\zeta,[A]}
=
\frac{
\sum_i
W_\zeta[A,U_i]\,
O[A^{U_i}]
\exp[-\beta F_i[A]]
}{
\sum_i
W_\zeta[A,U_i]\,
\exp[-\beta F_i[A]]
},
\label{app:weighted-average}
\end{equation}
with
\begin{equation}
W_\zeta[A,U_i]
=
s(i)
\frac{
\det({\cal M}[A^{U_i}]+\zeta{\bf 1})
}{
|\det{\cal M}[A^{U_i}]|
}.
\label{app:w-zeta}
\end{equation}
Provided that the denominator does not vanish on the minimizing set, the same localization gives
\begin{equation}
\lim_{\beta\to\infty}
\langle O\rangle_{\beta,\zeta,[A]}
=
\frac{
\sum_{i\in{\cal I}_*[A]}
W_\zeta[A,U_i]\,
O[A^{U_i}]
}{
\sum_{i\in{\cal I}_*[A]}
W_\zeta[A,U_i]
}.
\label{app:weighted-zero-temperature}
\end{equation}
Thus $W_\zeta$ can affect the residual measure among degenerate absolute minima, but it cannot change the zero-temperature support, as long as it does not remove all minimizing representatives.

This establishes the regulated mathematical basis for the FMR prescription used in the main text.
The essential point is that the $\beta\to\infty$ limit is always taken at fixed regulator, where the minimization problem is an ordinary compact-space problem.
Only after the absolute representative has been selected does one remove the Faddeev--Popov regulator, send the volume to infinity, and take the continuum limit.

\section{Finite-\texorpdfstring{$\beta$}{beta} saddle expansion}
\label{app:finite-beta-saddle}

In this section we derive the finite-$\beta$ expansion around an isolated absolute representative.
This is the technical origin of the $1/\beta$ correction quoted in the main text.

Let $U_*$ be an isolated absolute minimum of the Landau functional on a fixed gauge orbit.
A nearby gauge transformation can be parametrized as
\begin{equation}
U
=
U_*\exp(ig\theta),
\label{app:nearby-gauge-transformation}
\end{equation}
where $\theta=\theta^aT^a$ is small.
The gauge-transformed field is expanded as
\begin{equation}
A_\mu^U
=
A_\mu^{U_*}
+
D_\mu(A^{U_*})\theta
+
O(\theta^2).
\label{app:gauge-field-expansion}
\end{equation}
Since $U_*$ is a stationary point of $F_A[U]$, it satisfies the Landau condition
\begin{equation}
\partial_\mu A_\mu^{U_*}=0.
\label{app:landau-at-minimum}
\end{equation}
The second variation of the Landau functional is then governed by the Faddeev--Popov operator,
\begin{equation}
F_A[U]
=
F_A[U_*]
+
\frac{1}{2}
\int d^dx\,d^dy\,
\theta^a(x)
{\cal M}^{ab}(A^{U_*};x,y)
\theta^b(y)
+
O(\theta^3),
\label{app:landau-quadratic-expansion}
\end{equation}
with
\begin{equation}
{\cal M}^{ab}(A^{U_*})
=
-\partial_\mu D_\mu^{ab}(A^{U_*}).
\label{app:fp-at-absolute-representative}
\end{equation}
For an isolated minimum inside the FMR, the Hessian is positive on the nonzero modes after the global gauge zero modes are factored out.

Let $O[A^U]$ be a smooth orbit observable.
Near $U_*$ we write
\begin{eqnarray}
O[A^U]
&=&
O_*
+
\int d^dx\,
O^{(1),a}_*(x)\theta^a(x)
\nonumber\\
&&
+
\frac{1}{2}
\int d^dx\,d^dy\,
\theta^a(x)
O^{(2),ab}_*(x,y)
\theta^b(y)
+
O(\theta^3)
\nonumber\\[-0.7ex]
&&
\label{app:observable-expansion}
\end{eqnarray}
where
\begin{equation}
O_*
=
O[A^{U_*}].
\label{app:o-star}
\end{equation}
The finite-$\beta$ copy average is
\begin{equation}
\langle O\rangle_{\beta,[A]}
=
\frac{
\int{\cal D}\theta\,
O[A^{U_*\exp(ig\theta)}]\,
\exp[-\beta F_A[U_*\exp(ig\theta)]]
}{
\int{\cal D}\theta\,
\exp[-\beta F_A[U_*\exp(ig\theta)]]
},
\label{app:theta-average}
\end{equation}
where, for the saddle expansion, the integration is restricted to a neighborhood of the isolated minimum and the contribution from other copies is exponentially suppressed.

Using Eq.~\eqref{app:landau-quadratic-expansion}, the leading Gaussian measure is
\begin{equation}
d\mu_\beta(\theta)
=
\frac{
{\cal D}\theta\,
\exp\left[
-\frac{\beta}{2}
\int d^dx\,d^dy\,
\theta^a(x)
{\cal M}^{ab}(A^{U_*};x,y)
\theta^b(y)
\right]
}{
\int{\cal D}\theta\,
\exp\left[
-\frac{\beta}{2}
\int d^dx\,d^dy\,
\theta^a(x)
{\cal M}^{ab}(A^{U_*};x,y)
\theta^b(y)
\right]
}.
\label{app:gaussian-measure}
\end{equation}
The Gaussian contractions are
\begin{equation}
\langle\theta^a(x)\rangle_0=0,
\label{app:theta-one-point}
\end{equation}
and
\begin{equation}
\langle\theta^a(x)\theta^b(y)\rangle_0
=
\frac{1}{\beta}
\left[
{\cal M}^{-1}(A^{U_*})
\right]^{ab}(x,y).
\label{app:theta-two-point}
\end{equation}
Therefore, inserting Eq.~\eqref{app:observable-expansion} into Eq.~\eqref{app:theta-average}, one obtains
\begin{eqnarray}
\langle O\rangle_{\beta,[A]}
&=&
O[A^{U_*}]
+
\frac{1}{2\beta}
\int d^dx\,d^dy\,
\left[
{\cal M}^{-1}(A^{U_*})
\right]^{ab}(x,y)
\nonumber\\
&&
\hspace{2.0cm}\times
O^{(2),ba}_*(y,x)
+
O(\beta^{-2})
\nonumber\\[-0.7ex]
&&
\label{app:one-over-beta-integral}
\end{eqnarray}
Equivalently,
\begin{equation}
\langle O\rangle_{\beta,[A]}
=
O[A^{U_*}]
+
\frac{1}{2\beta}
\Tr\left[
{\cal M}^{-1}(A^{U_*})\,O^{(2)}_*
\right]
+
O(\beta^{-2}).
\label{app:one-over-beta-trace}
\end{equation}
This is the formula used in the main text.

Several comments are in order.
First, Eq.~\eqref{app:one-over-beta-trace} is an orbitwise statement.
It describes how the finite-temperature copy-weighted gauge approaches the FMR representative along a fixed orbit.
Second, the coefficient of the leading correction is controlled by the inverse Faddeev--Popov operator at the absolute representative.
Therefore, if the smallest nonzero eigenvalue of ${\cal M}(A^{U_*})$ is small, the approach to the FMR is slow.
Third, higher-order corrections involve higher variations of both $F_A[U]$ and $O[A^U]$ and generate the usual asymptotic Laplace expansion in powers of $1/\beta$.

To make the eigenvalue dependence explicit, let
\begin{equation}
{\cal M}(A^{U_*})\,\psi_n
=
\lambda_n\,\psi_n,
\qquad
\lambda_n>0
\label{app:fp-eigenvalue-equation}
\end{equation}
on the nonzero-mode subspace.
Then
\begin{equation}
{\cal M}^{-1}(A^{U_*})
=
\sum_n
\frac{
|\psi_n\rangle\langle\psi_n|
}{
\lambda_n
},
\label{app:fp-spectral-representation}
\end{equation}
and Eq.~\eqref{app:one-over-beta-trace} becomes
\begin{equation}
\langle O\rangle_{\beta,[A]}
=
O[A^{U_*}]
+
\frac{1}{2\beta}
\sum_n
\frac{
\langle\psi_n|O^{(2)}_*|\psi_n\rangle
}{
\lambda_n
}
+
O(\beta^{-2}).
\label{app:one-over-beta-spectral}
\end{equation}
This expression gives a practical diagnostic for lattice tests: copies whose absolute representatives lie close to the FMR boundary, where low FP eigenvalues occur, should display larger finite-$\beta$ corrections.

If determinant prefactors are included, the same saddle calculation applies to
\begin{equation}
\langle O\rangle_{\beta,\zeta,[A]}
=
\frac{
\int{\cal D}\theta\,
W_\zeta[A,U_*\exp(ig\theta)]\,
O[A^{U_*\exp(ig\theta)}]\,
e^{-\beta F_A[U_*\exp(ig\theta)]}
}{
\int{\cal D}\theta\,
W_\zeta[A,U_*\exp(ig\theta)]\,
e^{-\beta F_A[U_*\exp(ig\theta)]}
}.
\label{app:weighted-theta-average}
\end{equation}
For an isolated minimum with $W_\zeta[A,U_*]\neq0$, the leading term is unchanged,
\begin{equation}
\lim_{\beta\to\infty}
\langle O\rangle_{\beta,\zeta,[A]}
=
O[A^{U_*}],
\label{app:weighted-leading-term}
\end{equation}
whereas the $1/\beta$ coefficient receives additional contributions from the first and second variations of $\ln W_\zeta$.
Writing
\begin{equation}
L_\zeta[\theta]
=
\ln W_\zeta[A,U_*\exp(ig\theta)],
\label{app:log-w}
\end{equation}
one obtains schematically
\begin{eqnarray}
\langle O\rangle_{\beta,\zeta,[A]}
&=&
O[A^{U_*}]
+
\frac{1}{2\beta}
\Tr\left[
{\cal M}^{-1}(A^{U_*})\,O^{(2)}_*
\right]
\nonumber\\
&&
+
\frac{1}{\beta}
\int d^dx\,d^dy\,
\left[
{\cal M}^{-1}(A^{U_*})
\right]^{ab}(x,y)
\nonumber\\
&&
\hspace{1.4cm}\times
L^{(1),a}_{\zeta,*}(x)
O^{(1),b}_*(y)
+
O(\beta^{-2})
\nonumber\\[-0.7ex]
&&
\label{app:weighted-one-over-beta}
\end{eqnarray}
where $L^{(1)}_{\zeta,*}$ is the first variation of $\ln W_\zeta$ at the absolute representative.
Thus determinant factors can modify the subleading approach to the FMR, but not the support of the zero-temperature limit.

\section{Local finite-\texorpdfstring{$\beta$}{beta} replicated representation}
\label{app:local-finite-beta}

In this section we recall the local field-theory representation of the finite-$\beta$ copy-weighted gauge.
This is the local regulator whose zero-temperature limit defines the FMR prescription in the main text.
The exact FMR projection at $\beta=\infty$ is not itself local, because it implements a global minimization on the gauge orbit.

The starting point is the copy-weighted Landau-gauge average
\begin{equation}
\langle O[A]\rangle_{\beta,\zeta}
=
\frac{
\sum_i
O[A^{U_i}]
\,
\frac{
\det({\cal M}[A^{U_i}]+\zeta{\bf 1})
}{
|\det{\cal M}[A^{U_i}]|
}
\,
e^{-\beta F_A[U_i]}
}{
\sum_i
\frac{
\det({\cal M}[A^{U_i}]+\zeta{\bf 1})
}{
|\det{\cal M}[A^{U_i}]|
}
\,
e^{-\beta F_A[U_i]}
}.
\label{app:st-copy-average}
\end{equation}
The denominator is represented by the replica trick.
One introduces $n$ replicas and takes $n\to0$ at the end.
After singling out one replica as the ordinary Faddeev--Popov sector, the remaining replicas are represented by group-valued nonlinear sigma-model superfields
\begin{equation}
{\cal V}_k(x,\theta,\bar\theta)\in SU(N),
\qquad
k=2,\ldots,n,
\label{app:replica-superfields}
\end{equation}
satisfying
\begin{equation}
{\cal V}_k^\dagger{\cal V}_k={\bf 1}.
\label{app:replica-unitarity}
\end{equation}
The covariant derivative is
\begin{equation}
{\cal D}_\mu{\cal V}_k
=
\partial_\mu{\cal V}_k
+
ig\,{\cal V}_k A_\mu .
\label{app:replica-covariant-derivative}
\end{equation}

The local finite-$\beta$ action can be written as
\begin{eqnarray}
S_{\rm loc}^{(\beta,\zeta)}
&=&
S_{\rm YM}[A]
+
S_{\rm gf}^{(\beta,\zeta)}[A,c,\bar c,b]
\nonumber\\
&&
+
\sum_{k=2}^{n}
S_{\rm rep}[A,{\cal V}_k;\zeta]
\nonumber\\[-0.7ex]
&&
\label{app:local-beta-action}
\end{eqnarray}
where the singled replica gives the massive Faddeev--Popov/Curci--Ferrari type sector
\begin{equation}
S_{\rm gf}^{(\beta,\zeta)}
=
\int d^dx
\left[
\partial_\mu\bar c^aD_\mu^{ab}(A)c^b
+
\zeta\,\bar c^ac^a
+
ib^a\partial_\mu A_\mu^a
+
\frac{\beta}{2}A_\mu^aA_\mu^a
\right],
\label{app:singled-replica-action}
\end{equation}
and the remaining replicas are described by
\begin{eqnarray}
S_{\rm rep}[A,{\cal V}_k;\zeta]
&=&
\frac{1}{g^2}
\int d^dx\,d\theta\,d\bar\theta\,
{\rm tr}
\left[
({\cal D}_\mu{\cal V}_k)^\dagger
({\cal D}_\mu{\cal V}_k)
\right.
\nonumber\\
&&
\left.
\hspace{2.2cm}
+
2\zeta\,\bar\theta\theta\,
\partial_{\bar\theta}{\cal V}_k^\dagger
\partial_\theta{\cal V}_k
\right]
\nonumber\\[-0.7ex]
&&
\label{app:replica-super-action}
\end{eqnarray}
The expectation value at finite $\beta$ is then obtained as
\begin{equation}
\langle O\rangle_{\beta,\zeta}
=
\lim_{n\to0}
\frac{
\int{\cal D}\Phi\,
O[A]\,
e^{-S_{\rm loc}^{(\beta,\zeta)}}
}{
\int{\cal D}\Phi\,
e^{-S_{\rm loc}^{(\beta,\zeta)}}
},
\label{app:local-replica-expectation}
\end{equation}
where ${\cal D}\Phi$ denotes integration over the Yang--Mills field, the Faddeev--Popov fields, the Nakanishi--Lautrup field and the replica superfields.

At finite $\beta$, the quadratic part of the singled-replica gluon sector reads, in Landau gauge,
\begin{equation}
S_{AA}^{(2)}
=
\frac{1}{2}
\int\frac{d^dp}{(2\pi)^d}
A_\mu^a(-p)
\left[
(p^2+\beta)P_{\mu\nu}(p)
+
\frac{p^2}{\alpha}L_{\mu\nu}(p)
\right]
A_\nu^a(p),
\label{app:quadratic-gluon-beta}
\end{equation}
where
\begin{equation}
P_{\mu\nu}(p)
=
\delta_{\mu\nu}
-
\frac{p_\mu p_\nu}{p^2},
\qquad
L_{\mu\nu}(p)
=
\frac{p_\mu p_\nu}{p^2}.
\label{app:projectors}
\end{equation}
Consequently, the tree-level propagator of the local finite-$\beta$ regulator is
\begin{equation}
D_{\mu\nu}^{ab}(p)
=
\delta^{ab}
\left[
\frac{P_{\mu\nu}(p)}{p^2+\beta}
+
\alpha\,\frac{L_{\mu\nu}(p)}{p^2}
\right].
\label{app:tree-beta-propagator}
\end{equation}
This tree-level expression is not the main result of the paper; it only displays the local finite-$\beta$ regulator.
The nontrivial FMR content appears in the zero-temperature orbitwise localization and in the controlled $1/\beta$ expansion around the absolute representative.

For completeness, the ghost sector of the singled replica has the quadratic action
\begin{equation}
S_{\bar c c}^{(2)}
=
\int\frac{d^dp}{(2\pi)^d}\,
\bar c^a(-p)
(p^2+\zeta)c^a(p),
\label{app:ghost-quadratic}
\end{equation}
so that
\begin{equation}
G^{ab}(p)
=
\delta^{ab}
\frac{1}{p^2+\zeta}.
\label{app:ghost-propagator}
\end{equation}
The interaction vertices are the usual Faddeev--Popov vertices, supplemented by the replica nonlinear sigma-model interactions following from Eq.~\eqref{app:replica-super-action}.

The finite-$\beta$ local theory should be viewed as the continuum counterpart of a smooth copy-weighted gauge.
The FMR prescription is obtained by the ordered limit
\begin{equation}
\langle O\rangle_{\rm FMR}
=
\lim_{a\to0}
\lim_{V\to\infty}
\lim_{\zeta\to0^+}
\lim_{\beta\to\infty}
\lim_{n\to0}
\langle O\rangle_{n,\beta,\zeta}^{a,V}.
\label{app:localized-ordered-limit}
\end{equation}
The order of limits emphasizes that the zero-temperature projection is taken after the finite-$\beta$ local regulator has been defined.

If one wants to keep the relation with the first Gribov region explicit, the finite-$\beta$ local action can be supplemented by the BRST-invariant Gribov--Zwanziger block written in terms of $A_\mu^h$.
One then obtains
\begin{eqnarray}
S_{\rm loc}^{(\beta,\zeta,\gamma)}
&=&
S_{\rm YM}
+
S_{\rm gf}^{(\beta,\zeta)}
\nonumber\\
&&
+
\sum_{k=2}^{n}
S_{\rm rep}[A,{\cal V}_k;\zeta]
\nonumber\\
&&
-
\int d^dx\,
\bar\varphi_\mu^{ac}{\cal M}^{ab}(A^h)\varphi_\mu^{bc}
\nonumber\\
&&
+
\int d^dx\,
\bar\omega_\mu^{ac}{\cal M}^{ab}(A^h)\omega_\mu^{bc}
\nonumber\\
&&
-
g\gamma^2 f^{abc}
\int d^dx\,
(A^h)_\mu^a
(\varphi+\bar\varphi)_\mu^{bc}
\nonumber\\
&&
+
S_{\rm transv}[A^h]
+
S_{\rm ref}[A^h,\varphi,\omega]
\nonumber\\[-0.7ex]
&&
\label{app:local-beta-rgz-action}
\end{eqnarray}
Here $S_{\rm transv}$ enforces the transversality constraint
\begin{equation}
\partial_\mu A_\mu^h=0,
\label{app:ah-transversality}
\end{equation}
and $S_{\rm ref}$ denotes the usual refinement terms
\begin{equation}
S_{\rm ref}
=
\frac{m^2}{2}
\int d^dx\,
(A_\mu^{h,a})^2
-
M^2
\int d^dx\,
\left(
\bar\varphi_\mu^{ab}\varphi_\mu^{ab}
-
\bar\omega_\mu^{ab}\omega_\mu^{ab}
\right).
\label{app:refinement-terms}
\end{equation}
This combined finite-$\beta$ action displays the separation of mechanisms:
the horizon sector restricts the field measure toward the first Gribov region, while the zero-temperature copy-weighted sector selects the absolute representative on each orbit.

\section{Compatibility with the horizon restriction}
\label{app:horizon-compatibility}

We now spell out why the copy-weighted FMR localization is compatible with a horizon-based restriction to the first Gribov region.
The point is simple but important: the two mechanisms act on different variables.
The horizon sector changes the measure over gauge orbits, while the finite-$\beta$ copy weight changes the relative weight of representatives on a fixed orbit.

In the BRST-invariant formulation of the Gribov--Zwanziger construction, the horizon functional is written in terms of the transverse gauge-invariant field $A_\mu^h$,
\begin{equation}
H(A^h)
=
g^2
\int d^dx\,d^dy\,
f^{abc}A_\mu^{h,b}(x)
\left[
{\cal M}^{-1}(A^h)
\right]^{ad}(x,y)
f^{dec}A_\mu^{h,e}(y),
\label{app:horizon-functional}
\end{equation}
with
\begin{equation}
{\cal M}^{ab}(A^h)
=
-\partial_\mu D_\mu^{ab}(A^h).
\label{app:horizon-fp-operator}
\end{equation}
Since $A_\mu^h$ is gauge invariant, one has
\begin{equation}
(A^U)^h=A^h.
\label{app:ah-gauge-invariant}
\end{equation}
Therefore the horizon functional is constant along a fixed gauge orbit,
\begin{equation}
H((A^U)^h)
=
H(A^h).
\label{app:horizon-constant-orbit}
\end{equation}

Consider now the combined orbitwise weight
\begin{equation}
{\cal W}_{\beta,\gamma}[A,U_i]
=
\exp\left[
-\beta F_A[U_i]
-
\gamma^4H((A^{U_i})^h)
\right].
\label{app:combined-weight}
\end{equation}
Using Eq.~\eqref{app:horizon-constant-orbit}, this becomes
\begin{equation}
{\cal W}_{\beta,\gamma}[A,U_i]
=
e^{-\gamma^4H(A^h)}
e^{-\beta F_A[U_i]}.
\label{app:combined-weight-factorized}
\end{equation}
Hence the horizon factor cancels in any normalized copy average on the same orbit:
\begin{eqnarray}
\frac{
\sum_i O[A^{U_i}]
e^{-\beta F_A[U_i]-\gamma^4H(A^h)}
}{
\sum_i
e^{-\beta F_A[U_i]-\gamma^4H(A^h)}
}
&=&
\frac{
\sum_i O[A^{U_i}]
e^{-\beta F_A[U_i]}
}{
\sum_i
e^{-\beta F_A[U_i]}
}
\nonumber\\
&=&
\langle O\rangle_{\beta,[A]}.
\label{app:gamma-cancels-copy-average}
\end{eqnarray}
The same statement holds in the presence of the determinant prefactor $W_\zeta[A,U_i]$:
\begin{eqnarray}
&&
\frac{
\begin{gathered}
\sum_i W_\zeta[A,U_i]\,O[A^{U_i}]\\[-0.3ex]
\times e^{-\beta F_A[U_i]-\gamma^4H(A^h)}
\end{gathered}
}{
\begin{gathered}
\sum_i W_\zeta[A,U_i]\\[-0.3ex]
\times e^{-\beta F_A[U_i]-\gamma^4H(A^h)}
\end{gathered}
}
\nonumber\\
&=&
\frac{
\begin{gathered}
\sum_i W_\zeta[A,U_i]\,O[A^{U_i}]\\[-0.3ex]
\times e^{-\beta F_A[U_i]}
\end{gathered}
}{
\begin{gathered}
\sum_i W_\zeta[A,U_i]\\[-0.3ex]
\times e^{-\beta F_A[U_i]}
\end{gathered}
}
\nonumber\\[-0.7ex]
&&
\label{app:gamma-cancels-weighted-average}
\end{eqnarray}

This proves the orbitwise compatibility of the two operations.
The parameter $\beta$ controls the selection of representatives on a fixed orbit.
The parameter $\gamma$ controls the weighting of the orbit itself through the horizon sector.
Therefore, at the level of copy selection,
\begin{equation}
\gamma\neq0
\quad\text{does not modify}\quad
\lim_{\beta\to\infty}
\langle O\rangle_{\beta,[A]}.
\label{app:gamma-does-not-modify-beta-limit}
\end{equation}

Equivalently, one may write the combined expectation value schematically as
\begin{equation}
\langle O\rangle_{\beta,\gamma}
=
\frac{
\int{\cal D}A\,
e^{-S_{\rm YM}[A]-\gamma^4H(A^h)}
\,
\sum_i
O[A^{U_i}]
e^{-\beta F_A[U_i]}
}{
\int{\cal D}A\,
e^{-S_{\rm YM}[A]-\gamma^4H(A^h)}
\,
\sum_i
e^{-\beta F_A[U_i]}
}.
\label{app:combined-path-integral}
\end{equation}
The $\beta\to\infty$ limit acts on the copy sum at fixed orbit,
\begin{equation}
\sum_i
O[A^{U_i}]
e^{-\beta F_A[U_i]}
\quad
\underset{\beta\to\infty}{\longrightarrow}
\quad
e^{-\beta F_*[A]}
\sum_{i\in{\cal I}_*[A]}
O[A^{U_i}],
\label{app:beta-limit-copy-sum}
\end{equation}
whereas the horizon factor remains outside the copy selection.
Thus the combined prescription is
\begin{equation}
\langle O\rangle_{\FMR,\gamma}
=
\frac{
\int{\cal D}A\,
e^{-S_{\rm YM}[A]-\gamma^4H(A^h)}
\,
\displaystyle
\frac{1}{|{\cal I}_*[A]|}
\sum_{i\in{\cal I}_*[A]}
O[A^{U_i}]
}{
\int{\cal D}A\,
e^{-S_{\rm YM}[A]-\gamma^4H(A^h)}
}.
\label{app:fmr-with-horizon}
\end{equation}
For a unique absolute representative, this reduces to
\begin{equation}
\langle O\rangle_{\FMR,\gamma}
=
\frac{
\int{\cal D}A\,
e^{-S_{\rm YM}[A]-\gamma^4H(A^h)}
O[A^{U_*}]
}{
\int{\cal D}A\,
e^{-S_{\rm YM}[A]-\gamma^4H(A^h)}
}.
\label{app:fmr-with-horizon-unique}
\end{equation}

This makes precise the hierarchy discussed in the main text:
\begin{equation}
\text{Landau gauge}
\quad\longrightarrow\quad
\Omega
\quad\longrightarrow\quad
\Lambda.
\label{app:hierarchy}
\end{equation}
The restriction to $\Omega$ is controlled by the positivity of the Faddeev--Popov operator or, effectively, by the horizon sector.
The restriction to $\Lambda$ is controlled by the zero-temperature localization of the Landau functional on each gauge orbit.
These are complementary operations, not competing ones.

This also clarifies the issue of order of limits.
At the orbitwise level, the horizon factor is independent of the copy label.
Therefore the zero-temperature copy-selection limit and the insertion of the horizon weight commute as operations on the normalized copy average:
\begin{equation}
\lim_{\beta\to\infty}
\left[
\frac{
\sum_i O[A^{U_i}]
e^{-\beta F_A[U_i]-\gamma^4H(A^h)}
}{
\sum_i e^{-\beta F_A[U_i]-\gamma^4H(A^h)}
}
\right]
=
\lim_{\beta\to\infty}
\langle O\rangle_{\beta,[A]}.
\label{app:commutation-orbitwise}
\end{equation}
What remains nontrivial is the existence and stability of the subsequent infinite-volume and continuum limits of gauge-dependent correlation functions.
This is precisely why the main text defines the FMR through an ordered regulated limit rather than through an unregulated formal continuum expression.

\section{Lattice implementation and finite-\texorpdfstring{$\beta$}{beta} diagnostics}
\label{app:lattice-diagnostics}

The regulated construction has a direct lattice implementation.
For each gauge configuration $U_{x,\mu}$ one samples gauge transformations $g_x$ with probability
\begin{equation}
P_\beta[g|U]
=
\frac{
\exp[-\beta{\cal F}_U[g]]
}{
\int{\cal D}g\,\exp[-\beta{\cal F}_U[g]]
},
\label{app:lattice-p-beta}
\end{equation}
where
\begin{equation}
{\cal F}_U[g]
=
-\sum_{x,\mu}
{\rm Re\,Tr}\,
g_xU_{x,\mu}g^\dagger_{x+\hat\mu}.
\label{app:lattice-functional-diagnostics}
\end{equation}
The parameter $\beta$ controls how sharply the copy ensemble is concentrated around the best copy.
For $\beta=0$, all gauge transformations are weighted uniformly before imposing stationarity.
For large $\beta$, the measure concentrates on the absolute minima of ${\cal F}_U[g]$.

In practical gauge fixing, one does not integrate over all gauge transformations.
Instead, one generates a finite set of gauge-fixed copies,
\begin{equation}
\{g_1,g_2,\ldots,g_N\},
\label{app:finite-copy-set}
\end{equation}
obtained for example from random starts followed by overrelaxation or simulated annealing.
The finite-copy version of the copy-weighted average is
\begin{equation}
\langle O\rangle_{\beta,[U]}^{(N)}
=
\frac{
\sum_{r=1}^{N}
O[U^{g_r}]
\exp[-\beta{\cal F}_U[g_r]]
}{
\sum_{r=1}^{N}
\exp[-\beta{\cal F}_U[g_r]]
}.
\label{app:finite-copy-average}
\end{equation}
The best-copy estimator is recovered as
\begin{equation}
\lim_{\beta\to\infty}
\langle O\rangle_{\beta,[U]}^{(N)}
=
O[U^{g_{\rm bc}}],
\qquad
{\cal F}_U[g_{\rm bc}]
=
\min_{1\leq r\leq N}{\cal F}_U[g_r].
\label{app:finite-copy-best-copy}
\end{equation}
The absolute Landau gauge is approached by increasing both $\beta$ and the number of generated copies $N$.

The finite-$\beta$ prescription suggests a controlled extrapolation.
For each configuration, define
\begin{equation}
\Delta_r[U]
=
{\cal F}_U[g_r]-{\cal F}_U[g_{\rm bc}]
\geq0.
\label{app:finite-copy-delta}
\end{equation}
Then
\begin{equation}
\langle O\rangle_{\beta,[U]}^{(N)}
=
\frac{
O[U^{g_{\rm bc}}]
+
\sum_{r\neq{\rm bc}}
O[U^{g_r}]
\exp[-\beta\Delta_r[U]]
}{
1+
\sum_{r\neq{\rm bc}}
\exp[-\beta\Delta_r[U]]
}.
\label{app:finite-copy-delta-average}
\end{equation}
For large $\beta$, the leading correction is controlled by the copy closest to the best copy in functional value:
\begin{equation}
\langle O\rangle_{\beta,[U]}^{(N)}
=
O[U^{g_{\rm bc}}]
+
\sum_{r\neq{\rm bc}}
\left(
O[U^{g_r}]-O[U^{g_{\rm bc}}]
\right)
e^{-\beta\Delta_r[U]}
+
O(e^{-2\beta\Delta_{\rm min}}),
\label{app:finite-copy-large-beta}
\end{equation}
where
\begin{equation}
\Delta_{\rm min}
=
\min_{r\neq{\rm bc}}\Delta_r[U].
\label{app:finite-copy-min-gap}
\end{equation}
This formula is useful in numerical simulations because it separates two effects:
the finite-$\beta$ smoothing at fixed set of copies and the quality of the copy search itself.

For an ensemble of gauge fields, one may define the $\beta$-dependent gluon and ghost propagators by
\begin{eqnarray}
D_{\beta}(p)
&=&
\left\langle
\frac{1}{(N^2-1)(d-1)}
\sum_{a,\mu,\nu}
P_{\mu\nu}(p)\,
A_\mu^a(p)A_\nu^a(-p)
\right\rangle_{\beta}
\nonumber\\[-0.7ex]
&&
\label{app:lattice-gluon-beta}
\\[-0.2ex]
G_{\beta}(p)
&=&
\left\langle
\frac{1}{N^2-1}
\sum_a
\left[
{\cal M}^{-1}(A)
\right]^{aa}(p,-p)
\right\rangle_{\beta}
\nonumber\\[-0.7ex]
&&
\label{app:lattice-ghost-beta}
\end{eqnarray}
The FMR estimates are then obtained by the double extrapolation
\begin{eqnarray}
D_{\rm FMR}(p)
&=&
\lim_{N\to\infty}
\lim_{\beta\to\infty}
D_{\beta}^{(N)}(p),
\label{app:lattice-gluon-fmr}
\\
G_{\rm FMR}(p)
&=&
\lim_{N\to\infty}
\lim_{\beta\to\infty}
G_{\beta}^{(N)}(p).
\label{app:lattice-ghost-fmr}
\end{eqnarray}
In practice, the order of the two limits can be tested numerically by increasing $N$ at several large values of $\beta$.

The saddle expansion derived in Sec.~\ref{app:finite-beta-saddle} gives a sharper diagnostic.
For an isolated best copy, the leading continuum correction is
\begin{equation}
\delta O_\beta
\equiv
\langle O\rangle_{\beta,[A]}-O[A^{U_*}]
=
\frac{1}{2\beta}
\Tr\left[
{\cal M}^{-1}(A^{U_*})O_*^{(2)}
\right]
+
O(\beta^{-2}).
\label{app:lattice-delta-observable}
\end{equation}
Using the spectral decomposition of the Faddeev--Popov operator,
\begin{equation}
{\cal M}(A^{U_*})\psi_n=\lambda_n\psi_n,
\label{app:lattice-fp-spectrum}
\end{equation}
one obtains
\begin{equation}
\delta O_\beta
=
\frac{1}{2\beta}
\sum_n
\frac{
\langle\psi_n|O_*^{(2)}|\psi_n\rangle
}{
\lambda_n
}
+
O(\beta^{-2}).
\label{app:lattice-delta-spectrum}
\end{equation}
Therefore, the convergence toward the FMR should be slowest for configurations whose best copies have small low-lying Faddeev--Popov eigenvalues.

This leads to the following practical tests:
\begin{enumerate}
\item Compute $D_\beta(p)$ and $G_\beta(p)$ for several values of $\beta$ using the same set of gauge copies.
\item Extract the leading large-$\beta$ correction at fixed momentum.
\item Compare the fitted coefficient of $1/\beta$ or the exponential finite-copy correction with the lowest eigenvalues of ${\cal M}(A^{U_{\rm bc}})$.
\item Check whether the extrapolated values are stable as the number of generated copies $N$ is increased.
\item Repeat the analysis at different volumes and lattice spacings to test the stability of the ordered limit.
\end{enumerate}

The prediction specific to the copy-weighted FMR construction is not merely that the best-copy limit exists.
It is that the approach to this limit is controlled by the Faddeev--Popov spectrum at the absolute representative.
This gives an observable numerical target:
\begin{equation}
\text{small }\lambda_{\rm min}\left[{\cal M}(A^{U_{\rm bc}})\right]
\quad
\Longrightarrow
\quad
\text{slow convergence in }\beta.
\label{app:fp-eigenvalue-diagnostic}
\end{equation}
This diagnostic is independent of the detailed algorithm used to generate copies and can be tested directly in lattice simulations.

\section{Computational realization of the finite-temperature orbit measure}
\label{app:computational-realization}

This section describes a small numerical proof of concept for the regulated
orbit measure. The purpose is not to claim that the absolute minimum can be
certified on large lattices, but to demonstrate that the finite-$\beta$ family
and the proposed FP spectral diagnostic lead to explicit computational
procedures.

\subsection{Local relaxation and simulated annealing}

For $SU(2)$, links and gauge transformations were represented as unit
quaternions. At each site, the local gauge transformation that maximizes the
sum of the real traces of the eight incident links can be obtained exactly.
Successive local updates, optionally raised to an overrelaxation power
$1<\omega<2$, were applied until the lattice Landau residual was numerically
small. Simulated annealing used local random $SU(2)$ transformations accepted
with the finite-temperature weight and was followed by the same deterministic
relaxation.

\subsection{Population annealing}

For each gauge orbit, a population of $R$ randomly transformed representatives
was initialized at $\beta_0=0$. When increasing the inverse temperature from
$\beta_k$ to $\beta_{k+1}$, replica $r$ received the incremental weight
\begin{equation}
 w_r^{(k)}=
 \exp\left[-(\beta_{k+1}-\beta_k)
 {\cal F}_U[g_r]\right].
\label{app:pa-weight}
\end{equation}
The population was resampled according to the normalized weights and then
rejuvenated by local Metropolis sweeps preserving
$P_{\beta_{k+1}}[g|U]$. This is the direct sequential-Monte-Carlo realization
of the copy-weighted measure. The effective population size
\begin{equation}
 R_{\rm eff}^{(k)}=
 \frac{\left(\sum_r w_r^{(k)}\right)^2}
 {\sum_r\left(w_r^{(k)}\right)^2}
\label{app:pa-ess}
\end{equation}
was recorded as a diagnostic. The final population was locally relaxed before
comparing extrema.

\subsection{Evolutionary neural warm start}

The proof of concept also used a small group-preserving neural controller. Its
inputs were local invariant combinations of the lattice divergence and a
parallel-transported neighbor average. Its scalar outputs multiplied three
adjoint-vector directions: the local divergence, a covariant Laplacian-like
direction, and their cross product. The update was exponentiated in the
$SU(2)$ Lie algebra, guaranteeing an $SU(2)$-valued transformation.

The controller contained only eighteen parameters and was optimized by a
cross-entropy evolutionary strategy with an orbitwise soft-min objective. It
was used only to generate an initial gauge transformation. The resulting copy
was then relaxed by the same deterministic solver used for all other methods.
All neural updates were evaluated in double precision in the benchmark, so the
final links remained on the original gauge orbit up to numerical precision.

\subsection{Faddeev--Popov-guided basin hopping}

At a locally converged representative $U^{g_*}$, the lattice FP Hessian was
constructed and its lowest nontrivial eigenpairs were evaluated,
\begin{equation}
 {\cal M}(U^{g_*})\psi_n=\lambda_n\psi_n.
\end{equation}
Collective trial transformations were generated as
\begin{equation}
 h(x)=\exp\left[i\theta^a(x)T^a\right],
 \qquad
 \theta^a(x)=
 \sum_{n=1}^{N_{\rm low}}
 c_n\psi_n^a(x),
\end{equation}
with
\begin{equation}
 c_n=\sigma\,
 \frac{\xi_n}{\sqrt{\lambda_n+\mu^2}}.
\label{app:fp-coefficients}
\end{equation}
The field $\theta$ was normalized to a prescribed maximum site angle and three
amplitudes were scanned. Each displaced copy was subsequently relaxed to a
stationary point. An isotropic control used independent Gaussian site fields
with the same maximum-angle budget and the same number of candidates.

\subsection{Exploratory ensemble and benchmark}

The benchmark used twelve stored configurations from a small $SU(2)$
Wilson-action ensemble on a $4^4$ lattice at $\beta_{\rm W}=1.2$. The ensemble
was generated only for this proof of concept; production-level thermalization,
autocorrelation and continuum analyses were not attempted. The low-budget
methods used four random starts. Population annealing used $R=24$ replicas.
The collective methods used eight candidates at each of three amplitudes. A
thirty-two-start overrelaxation search was used as a finite reference.

The benchmark-best value for each orbit was defined as the lowest Landau loss
obtained by any included method. Consequently, a reported hit means agreement
with the best value found in this calculation, not a certificate of membership
in the exact FMR.

\begin{figure}[t]
\centering
\includegraphics[width=\columnwidth]{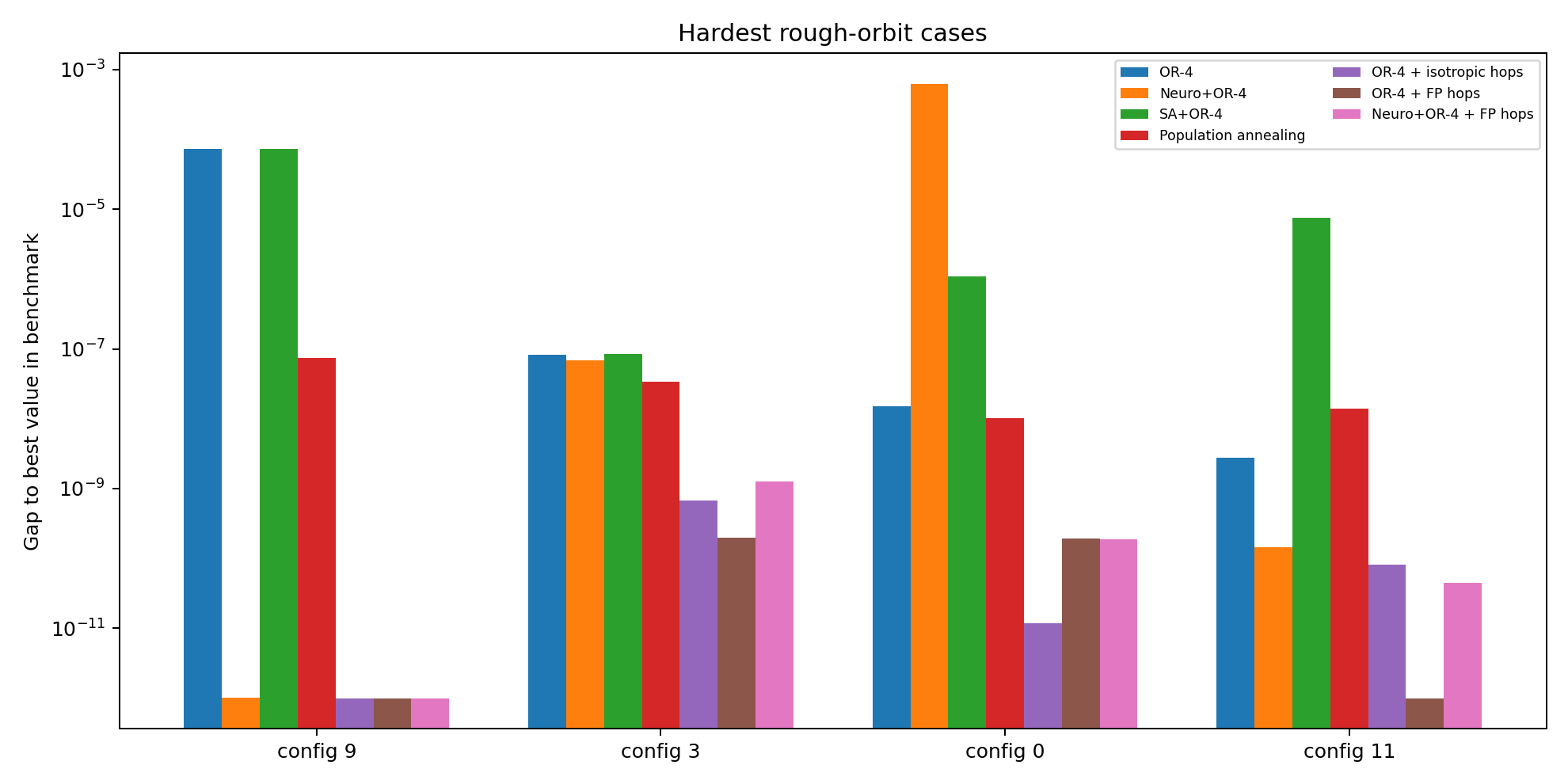}
\caption{Gap to the lowest value found in the benchmark for the four most
difficult configurations according to the four-start overrelaxation result.
The collective-hop methods recover lower basins missed by the low-budget local
and simulated-annealing searches.}
\label{fig:hard-orbits}
\end{figure}

\begin{figure}[t]
\centering
\includegraphics[width=\columnwidth]{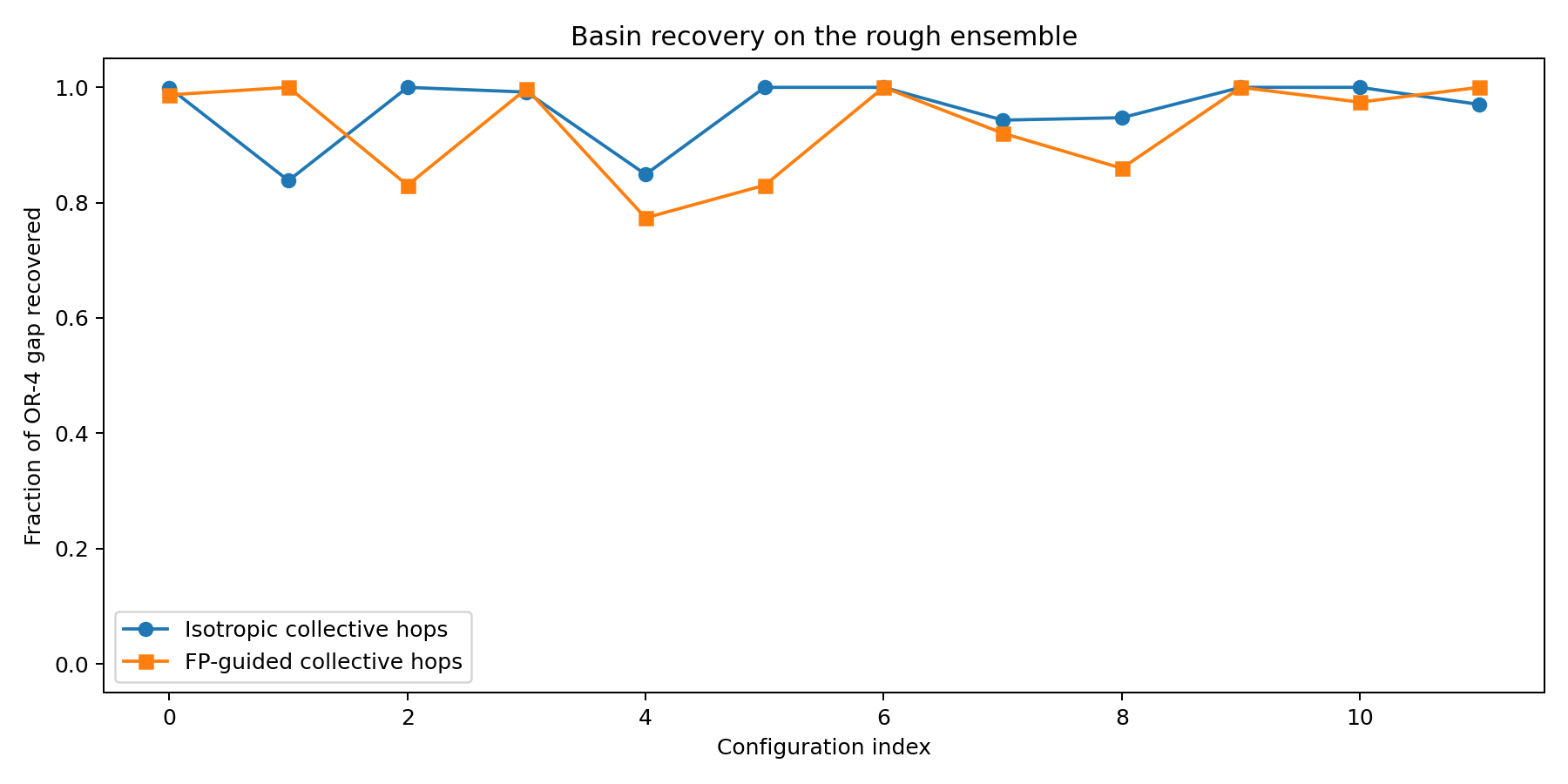}
\caption{Fraction of the four-start overrelaxation gap recovered by isotropic
and FP-guided collective hopping. Both methods are successful on the trapped
orbits of this small ensemble, but no systematic advantage of the FP-guided
proposal is resolved at this volume.}
\label{fig:basin-recovery}
\end{figure}

The aggregated results are summarized in the main text. Four-start overrelaxation agreed with the benchmark-best value
within $10^{-7}$ on eleven of twelve configurations, but on one orbit it was
trapped approximately $7.3\times10^{-5}$ above the lower basin. Population
annealing reached the $10^{-7}$ tolerance on all twelve orbits. Both collective
proposal families also reached that tolerance on all twelve orbits, with mean
gaps of order $10^{-11}$ in the present normalization.

The result supports the practical content of the zero-temperature formulation:
one may sample the finite-$\beta$ orbit measure directly and may use the FP
spectrum to construct collective moves. The present test does not show that
spectral guidance outperforms generic collective hopping, nor that the neural
warm start improves every orbit. Establishing either point requires larger
volumes, independent ensembles, tuned baselines and measurements of
$D_\beta(p)$ and $G_\beta(p)$.

\section{Extension to full QCD and FMR-resolved generating functional}
\label{sec:qcd-extension}

The construction extends directly to full QCD and yields a regulated definition
of the gauge-fixed QCD generating functional in absolute Landau gauge.
This should not be confused with a closed-form solution of QCD. A solution of
QCD in the strong sense would require the explicit evaluation of the full
generating functional, including the hadron spectrum, the mass gap,
confinement, and all correlation functions. The result obtained here is
different: it is a closed regulated prescription for the nonperturbative QCD
functional integral after the residual Landau-gauge Gribov ambiguity has been
removed by the FMR projection.

At fixed ultraviolet and infrared regulator, consider the lattice QCD measure
\begin{equation}
{\cal D}U_{\rm lat}\,
{\cal D}\bar\psi\,
{\cal D}\psi\,
\exp\left[
-S_{\rm YM}[U_{\rm lat}]
-S_{\rm q}[U_{\rm lat},\bar\psi,\psi]
\right],
\label{qcd-measure}
\end{equation}
where $S_{\rm q}$ denotes the quark action. Since the quark action is gauge
invariant,
\begin{equation}
S_{\rm q}[U_{\rm lat}^g,\bar\psi^g,\psi^g]
=
S_{\rm q}[U_{\rm lat},\bar\psi,\psi],
\label{quark-gauge-invariance}
\end{equation}
the orbitwise copy selection is still entirely governed by the Landau
functional ${\cal F}_{U_{\rm lat}}[g]$.

For each gauge orbit we introduce the orbitwise normalization
\begin{equation}
{\cal N}_{\beta,\zeta}[U_{\rm lat}]
=
\int{\cal D}g\,
W_\zeta[U_{\rm lat},g]\,
e^{-\beta{\cal F}_{U_{\rm lat}}[g]}.
\label{qcd-orbit-normalization}
\end{equation}
For a possibly gauge-dependent QCD observable
$O[U,\bar\psi,\psi]$, the regulated copy-weighted expectation value is then
defined as
\begin{eqnarray}
\langle O\rangle_{\beta,\zeta,{\rm QCD}}^{a,V}
&=&
\frac{1}{Z_{\rm QCD}^{a,V}}
\int{\cal D}U_{\rm lat}\,
{\cal D}\bar\psi\,
{\cal D}\psi\,
e^{-S_{\rm YM}[U_{\rm lat}]
-S_{\rm q}[U_{\rm lat},\bar\psi,\psi]}
\nonumber\\
&&\times
\frac{
\int{\cal D}g\,
W_\zeta[U_{\rm lat},g]\,
O[U_{\rm lat}^g,\bar\psi^g,\psi^g]\,
e^{-\beta{\cal F}_{U_{\rm lat}}[g]}
}{
{\cal N}_{\beta,\zeta}[U_{\rm lat}]
}
\nonumber\\[-0.7ex]
&&
\label{qcd-copy-weighted-expectation}
\end{eqnarray}
with
\begin{equation}
Z_{\rm QCD}^{a,V}
=
\int{\cal D}U_{\rm lat}\,
{\cal D}\bar\psi\,
{\cal D}\psi\,
e^{-S_{\rm YM}[U_{\rm lat}]
-S_{\rm q}[U_{\rm lat},\bar\psi,\psi]}.
\label{qcd-partition-function}
\end{equation}
The QCD expectation value in absolute Landau gauge, or equivalently in the
FMR prescription, is then defined by the ordered limit
\begin{eqnarray}
&&
\boxed{
\begin{gathered}
\langle O\rangle_{{\rm QCD},\FMR}
=
\lim_{a\to0}
\lim_{V\to\infty}
\lim_{\zeta\to0^+}
\lim_{\beta\to\infty}
\langle O\rangle_{\beta,\zeta,{\rm QCD}}^{a,V}
\end{gathered}
}
\nonumber\\[-0.7ex]
&&
\label{qcd-fmr-limit}
\end{eqnarray}
This is the precise sense in which the present construction defines the
nonperturbative QCD vacuum in absolute Landau gauge. It is not an explicit
solution of QCD, but a regulated FMR prescription for its gauge-fixed
functional integral.

The corresponding FMR-resolved generating functional is
\begin{eqnarray}
Z_{\beta,\zeta}^{a,V}[J,\eta,\bar\eta]
&=&
\frac{1}{Z_{\rm QCD}^{a,V}}
\int{\cal D}U_{\rm lat}\,
{\cal D}\bar\psi\,
{\cal D}\psi
\nonumber\\
&&\times
e^{-S_{\rm YM}[U_{\rm lat}]
-S_{\rm q}[U_{\rm lat},\bar\psi,\psi]}
\nonumber\\
&&\times
\frac{1}{{\cal N}_{\beta,\zeta}[U_{\rm lat}]}
\int{\cal D}g\,
W_\zeta[U_{\rm lat},g]
\nonumber\\
&&\times
e^{-\beta{\cal F}_{U_{\rm lat}}[g]}
\exp\left[
J\cdot A^g
+
\bar\eta\,\psi^g
+
\bar\psi^g\,\eta
\right]
\nonumber\\[-0.7ex]
&&
\label{qcd-beta-zeta-generating-functional}
\end{eqnarray}
Here $J$ is a source for the gauge field, while $\eta$ and $\bar\eta$ are
Grassmann sources for the quark fields. The FMR-resolved QCD generating
functional is defined by
\begin{equation}
\boxed{
Z_{{\rm QCD},\FMR}[J,\eta,\bar\eta]
=
\lim_{a\to0}
\lim_{V\to\infty}
\lim_{\zeta\to0^+}
\lim_{\beta\to\infty}
Z_{\beta,\zeta}^{a,V}[J,\eta,\bar\eta].
}
\label{qcd-fmr-generating-functional}
\end{equation}
Gauge-fixed QCD Green functions in the FMR prescription are obtained by
functional differentiation. For example,
\begin{eqnarray}
\left\langle
A_{\mu_1}^{a_1}(x_1)
\cdots
A_{\mu_n}^{a_n}(x_n)
\right\rangle_{{\rm QCD},\FMR}
&=&
\left.
\frac{
\delta^n Z_{{\rm QCD},\FMR}[J,0,0]
}{
\delta J_{\mu_1}^{a_1}(x_1)
\cdots
\delta J_{\mu_n}^{a_n}(x_n)
}
\right|_{J=0}
\nonumber\\[-0.7ex]
&&
\label{qcd-fmr-gluon-correlators}
\end{eqnarray}
Similarly, quark and mixed quark--gluon Green functions are obtained by
differentiating with respect to $\eta$, $\bar\eta$ and $J$.

In formal continuum notation, the same object may be represented as
\begin{eqnarray}
Z_{{\rm QCD},\FMR}[J,\eta,\bar\eta]
&=&
\int_{\Lambda}
{\cal D}A\,
{\cal D}\bar\psi\,
{\cal D}\psi\,
\det{\cal M}(A)
\nonumber\\
&&\times
e^{-S_{\rm YM}[A]-S_{\rm q}[A,\bar\psi,\psi]}
\nonumber\\
&&\times
\exp\left[
J\cdot A
+
\bar\eta\,\psi
+
\bar\psi\,\eta
\right]
\nonumber\\[-0.7ex]
&&
\label{formal-qcd-fmr-functional}
\end{eqnarray}
where $\Lambda$ denotes the Fundamental Modular Region. Equation
\eqref{formal-qcd-fmr-functional} should be understood only as a compact
shorthand for the regulated definition
\eqref{qcd-fmr-generating-functional}; the restriction to $\Lambda$ is
implemented by the zero-temperature copy-weighted limit.

After integrating out the quark fields one obtains, formally,
\begin{eqnarray}
Z_{{\rm QCD},\FMR}[J]
&=&
\int_{\Lambda}
{\cal D}A\,
\det{\cal M}(A)\,
\det\left(\slashed{D}[A]+m\right)
\nonumber\\
&&\times
\exp\left[
-S_{\rm YM}[A]
+
J\cdot A
\right]
\nonumber\\[-0.7ex]
&&
\label{formal-qcd-fmr-quark-determinant}
\end{eqnarray}
again with the understanding that the FMR restriction is defined by the
ordered regulated limit.

For gauge-invariant observables, the orbitwise copy average is trivial. If
\begin{equation}
O[U^g,\bar\psi^g,\psi^g]
=
O[U,\bar\psi,\psi],
\label{qcd-gauge}
\end{equation}

\section{Hamiltonian interpretation: vacuum wave functional on the FMR}
\label{app:hamiltonian-fmr}

The zero-temperature construction also admits a Hamiltonian interpretation.
This interpretation is useful because it makes explicit in what sense the
prescription defines a gauge-fixed representation of the nonperturbative
vacuum, without claiming an explicit solution of the Yang--Mills or QCD
Hamiltonian.

Consider the Schrödinger representation on a fixed time slice. The
configuration variable is the spatial gauge field $A_i^a(\mathbf{x})$.
Physical wave functionals obey Gauss' law,
\begin{equation}
\widehat{\cal G}^a(\mathbf{x})\,\Psi_{\rm phys}[A]=0,
\label{app:gauss-law-constraint}
\end{equation}
where, in pure Yang--Mills theory,
\begin{equation}
\widehat{\cal G}^a(\mathbf{x})
=
\left(D_i^{ab}(A)\widehat E_i^b\right)(\mathbf{x}).
\label{app:gauss-law-pure-ym}
\end{equation}
In full QCD, the matter color charge is included in the Gauss-law generator.
As a consequence, physical states are invariant under time-independent gauge
transformations,
\begin{equation}
\Psi_{\rm phys}[A^g]=\Psi_{\rm phys}[A],
\label{app:physical-wavefunctional-gauge-invariant}
\end{equation}
or, in the presence of quarks,
\begin{equation}
\Psi_{\rm phys}[A^g,\psi^g]
=
\Psi_{\rm phys}[A,\psi].
\label{app:physical-wavefunctional-qcd-gauge-invariant}
\end{equation}

The vacuum wave functional $\Psi_0$ is therefore constant on gauge orbits.
However, gauge-dependent quantities require a choice of representative on each
orbit. The FMR prescription provides precisely such a choice. On a fixed time
slice, define the spatial Landau functional
\begin{equation}
{\cal F}_{A}[g]
=
\frac{1}{2}
\int d^{d_s}x\,
A_i^{g,a}(\mathbf{x})A_i^{g,a}(\mathbf{x}),
\label{app:hamiltonian-landau-functional}
\end{equation}
where $d_s$ is the number of spatial dimensions. Its stationary points obey
\begin{equation}
\partial_i A_i^g=0,
\label{app:hamiltonian-landau-condition}
\end{equation}
and its Hessian at a stationary point is the spatial Faddeev--Popov operator
\begin{equation}
{\cal M}_{\rm H}(A^g)
=
-\partial_iD_i(A^g).
\label{app:hamiltonian-fp-operator}
\end{equation}

At finite regulator, the copy-weighted Hamiltonian expectation value of a
possibly gauge-dependent observable $O[A]$ may be written orbitwise as
\begin{eqnarray}
\langle O\rangle_{\beta,\zeta}^{\rm H}
&=&
\frac{1}{Z_{\rm H}}
\int{\cal D}A\,
|\Psi_0[A]|^2
\,
\frac{
\int{\cal D}g\,
W_\zeta[A,g]\,
O[A^g]\,
e^{-\beta{\cal F}_{A}[g]}
}{
\int{\cal D}g\,
W_\zeta[A,g]\,
e^{-\beta{\cal F}_{A}[g]}
}
\nonumber\\[-0.7ex]
&&
\label{app:hamiltonian-copy-weighted-average}
\end{eqnarray}
Here
\begin{equation}
Z_{\rm H}
=
\int{\cal D}A\,
|\Psi_0[A]|^2 .
\label{app:hamiltonian-normalization}
\end{equation}
Equation~\eqref{app:hamiltonian-copy-weighted-average} is formal in the
continuum and should be understood with the same ultraviolet and infrared
regulators used in the main construction. The normalized orbit average removes
the irrelevant gauge-volume factor and makes the copy selection explicit.

Let $g_*[A]$ denote an absolute minimizer of the spatial Landau functional on
the orbit of $A$,
\begin{equation}
{\cal F}_{A}[g_*]
=
\min_g{\cal F}_{A}[g].
\label{app:hamiltonian-best-copy}
\end{equation}
For a unique absolute representative, the zero-temperature limit gives
\begin{equation}
\lim_{\beta\to\infty}
\langle O\rangle_{\beta,\zeta}^{\rm H}
=
\frac{1}{Z_{\rm H}}
\int{\cal D}A\,
|\Psi_0[A]|^2\,
O[A^{g_*[A]}].
\label{app:hamiltonian-fmr-limit-orbit}
\end{equation}
If several absolute representatives exist, the limiting prescription gives the
corresponding residual average over the degenerate minimizing set, as in the
path-integral construction.

Thus the vacuum wave functional can be represented on the FMR by defining
\begin{equation}
A_\Lambda
=
A^{g_*[A]},
\qquad
A_\Lambda\in\Lambda,
\label{app:hamiltonian-fmr-representative}
\end{equation}
and
\begin{equation}
\Psi_{0,\Lambda}[A_\Lambda]
=
\Psi_0[A].
\label{app:fmr-vacuum-wavefunctional}
\end{equation}
This is well defined because $\Psi_0$ is constant on gauge orbits. In formal
FMR notation, Eq.~\eqref{app:hamiltonian-fmr-limit-orbit} may be rewritten as
\begin{equation}
\langle O\rangle_{\FMR}^{\rm H}
=
\frac{
\int_{\Lambda}{\cal D}A\,
J_\Lambda[A]\,
|\Psi_{0,\Lambda}[A]|^2\,
O[A]
}{
\int_{\Lambda}{\cal D}A\,
J_\Lambda[A]\,
|\Psi_{0,\Lambda}[A]|^2
}.
\label{app:hamiltonian-fmr-expectation}
\end{equation}
Here $J_\Lambda[A]$ denotes the induced measure on the FMR. Equation
\eqref{app:hamiltonian-fmr-expectation} is only a compact formal notation; the
regulated definition is the zero-temperature copy-weighted limit
\eqref{app:hamiltonian-copy-weighted-average}.

For gauge-invariant observables, the copy average is trivial. If
\begin{equation}
O[A^g]=O[A],
\label{app:hamiltonian-gauge-invariant-observable}
\end{equation}
then
\begin{equation}
\langle O\rangle_{\beta,\zeta}^{\rm H}
=
\frac{1}{Z_{\rm H}}
\int{\cal D}A\,
|\Psi_0[A]|^2\,
O[A],
\label{app:hamiltonian-gauge-invariant-unchanged}
\end{equation}
independently of $\beta$. Therefore, the FMR projection does not modify
gauge-invariant physics. Its role is to define gauge-dependent vacuum
correlators, such as equal-time gluon or quark Green functions, in a
representative free of residual Landau-gauge Gribov ambiguity.

The finite-$\beta$ approach is again controlled by the Faddeev--Popov spectrum.
For an isolated absolute representative $A_\Lambda=A^{g_*}$, the orbitwise
saddle expansion gives
\begin{equation}
\langle O\rangle_{\beta,[A]}^{\rm H}
=
O[A_\Lambda]
+
\frac{1}{2\beta}
\Tr\left[
{\cal M}_{\rm H}^{-1}(A_\Lambda)\,
O^{(2)}_\Lambda
\right]
+
O(\beta^{-2}),
\label{app:hamiltonian-one-over-beta}
\end{equation}
where $O^{(2)}_\Lambda$ is the second variation of the observable along the
gauge orbit at the absolute representative. In spectral form,
\begin{equation}
{\cal M}_{\rm H}(A_\Lambda)\psi_n
=
\lambda_n\psi_n,
\qquad
\lambda_n>0,
\label{app:hamiltonian-fp-spectrum}
\end{equation}
and
\begin{equation}
\langle O\rangle_{\beta,[A]}^{\rm H}
=
O[A_\Lambda]
+
\frac{1}{2\beta}
\sum_n
\frac{
\langle\psi_n|O^{(2)}_\Lambda|\psi_n\rangle
}{
\lambda_n
}
+
O(\beta^{-2}).
\label{app:hamiltonian-one-over-beta-spectrum}
\end{equation}
Thus, also in the Hamiltonian language, the convergence toward the FMR
representative is governed by the low-lying spectrum of the Faddeev--Popov
operator at the absolute minimum.

The Hamiltonian interpretation therefore gives an equivalent view of the main
construction: the physical vacuum wave functional remains gauge invariant, but
its gauge-fixed representation is obtained by selecting the absolute Landau
representative on each orbit through a regulated zero-temperature limit. This
provides a well-defined FMR representation of the vacuum functional without
requiring an explicit parametrization of the FMR boundary or an explicit
solution of the Hamiltonian eigenvalue problem.

\section{Ordered limits and summary of the prescription}
\label{app:ordered-limits}

The construction used in the main text is a regulated prescription.
The role of the regulator is not cosmetic: at fixed ultraviolet cutoff and finite volume, the orbitwise minimization problem is well defined, while in the formal continuum the geometry of the FMR is global and analytically inaccessible.
The complete ordered definition is
\begin{equation}
\langle O\rangle_{\rm FMR}
=
\lim_{a\to0}
\lim_{V\to\infty}
\lim_{\zeta\to0^+}
\lim_{\beta\to\infty}
\langle O\rangle_{\beta,\zeta}^{a,V}.
\label{app:complete-ordered-limit}
\end{equation}
Here $\langle O\rangle_{\beta,\zeta}^{a,V}$ denotes the finite-cutoff, finite-volume copy-weighted expectation value,
\begin{equation}
\langle O\rangle_{\beta,\zeta}^{a,V}
=
\frac{
\int{\cal D}U_{\rm lat}\,e^{-S_{\rm YM}[U_{\rm lat}]}
\int{\cal D}g\,
W_\zeta[U_{\rm lat},g]\,
O[U_{\rm lat}^g]\,
e^{-\beta{\cal F}_{U_{\rm lat}}[g]}
}{
\int{\cal D}U_{\rm lat}\,e^{-S_{\rm YM}[U_{\rm lat}]}
\int{\cal D}g\,
W_\zeta[U_{\rm lat},g]\,
e^{-\beta{\cal F}_{U_{\rm lat}}[g]}
}.
\label{app:regulated-expectation-final}
\end{equation}
The determinant prefactor is
\begin{equation}
W_\zeta[U_{\rm lat},g]
=
s(g)
\frac{
\det\left({\cal M}[U_{\rm lat}^{g}]+\zeta{\bf 1}\right)
}{
\left|\det{\cal M}[U_{\rm lat}^{g}]\right|
},
\qquad
s(g)=\mathrm{sign}\det{\cal M}[U_{\rm lat}^{g}].
\label{app:wzeta-lattice-final}
\end{equation}
The order of limits in Eq.~\eqref{app:complete-ordered-limit} has the following meaning.
First,
\begin{equation}
\beta\to\infty
\label{app:beta-limit-meaning}
\end{equation}
performs the absolute-minimum selection on each orbit at fixed regulator.
This is the step that implements the FMR.
Second,
\begin{equation}
\zeta\to0^+
\label{app:zeta-limit-meaning}
\end{equation}
removes the FP regulator after the minimizing representatives have been selected.
Third,
\begin{equation}
V\to\infty,
\qquad
a\to0
\label{app:volume-continuum-limits}
\end{equation}
define the infinite-volume and continuum limits of the resulting gauge-fixed correlation functions.
The finite-$\beta$ local replicated action represents the regulator before the FMR projection is taken.
Schematically,
\begin{equation}
S_{\rm loc}^{(\beta,\zeta)}
=
S_{\rm YM}
+
S_{\rm gf}^{(\beta,\zeta)}
+
\sum_{k=2}^{n}S_{\rm rep}[A,{\cal V}_k;\zeta],
\qquad
n\to0.
\label{app:finite-beta-local-summary}
\end{equation}
At finite $\beta$, this is a local continuum field theory.
The strict $\beta=\infty$ limit is not expected to be described by a simple local action, because it encodes an absolute minimization over the full gauge orbit.
Thus the locality statement is
\begin{eqnarray}
\text{finite }\beta
&:&
\text{local replicated regulator},
\nonumber\\
\beta=\infty
&:&
\text{global FMR projection}
\nonumber\\[-0.7ex]
&&
\label{app:locality-summary}
\end{eqnarray}
The controlled approach to the FMR is governed by the saddle expansion
\begin{equation}
\langle O\rangle_{\beta,[A]}
=
O[A^{U_*}]
+
\frac{1}{2\beta}
\Tr\left[
{\cal M}^{-1}(A^{U_*})\,O^{(2)}_*
\right]
+
O(\beta^{-2}),
\label{app:controlled-approach-summary}
\end{equation}
valid for an isolated absolute minimum.
This formula shows that the leading correction is controlled by the inverse Faddeev--Popov operator at the absolute representative.
Consequently, the low-lying FP spectrum controls the rate at which finite-$\beta$ gauges approach the FMR.
The compatibility with horizon-based approaches follows from the gauge invariance of $A^h$.
Since
\begin{equation}
H((A^U)^h)=H(A^h),
\label{app:horizon-summary}
\end{equation}
the horizon factor is constant along each gauge orbit and cancels in normalized copy averages.
Therefore, the two mechanisms act on different structures:
\begin{equation}
\gamma\neq0
\quad\Longrightarrow\quad
\text{restriction of the orbit measure toward }\Omega,
\label{app:gamma-summary}
\end{equation}
whereas
\begin{equation}
\beta\to\infty
\quad\Longrightarrow\quad
\text{absolute-minimum selection on each orbit, i.e. }\Lambda.
\label{app:beta-summary}
\end{equation}
This yields the hierarchy
\begin{equation}
\text{Landau gauge}
\quad\longrightarrow\quad
\Omega
\quad\longrightarrow\quad
\Lambda.
\label{app:hierarchy-summary-final}
\end{equation}
The practical lattice consequence is that the FMR limit can be approached by a finite-temperature copy ensemble and tested against best-copy or simulated-annealing gauge fixing.
The predicted diagnostic is
\begin{eqnarray}
\delta O_\beta
&=&
\langle O\rangle_{\beta,[A]}-O[A^{U_*}]
=
O(\beta^{-1}),
\nonumber\\
&&
\delta O_\beta
\ \text{enhanced by small FP eigenvalues}
\nonumber\\[-0.7ex]
&&
\label{app:lattice-diagnostic-summary}
\end{eqnarray}
Thus the construction provides more than the formal statement that absolute minima exist at finite regulator.
It gives a controlled expansion around the FMR representative and a direct observable criterion for numerical tests.

\bibliographystyle{elsarticle-num}
\bibliography{references}
\end{document}